\documentclass[aps,twocolumn,showpacs,superscriptaddress,amsmath,amssymb]{revtex4}
\usepackage{graphicx}
\usepackage[latin1]{inputenc}
\usepackage{dcolumn}
\usepackage{bm}
\usepackage{color}

\begin{document}

\title{Semiclassical magneto-transport in graphene $n$-$p$ junctions}  
\author{Pierre Carmier} 
\affiliation{Univ.\ Paris-Sud, LPTMS UMR 8626, 91405 Orsay Cedex, France}
\affiliation{CEA-INAC/UJF Grenoble 1, SPSMS UMR-E 9001, 38054 Grenoble Cedex 9, France}
\author{Caio Lewenkopf} 
\affiliation{Instituto de F\'{\i}sica, Universidade Federal
  Fluminense, 24210-346 Niter\'oi RJ, Brazil}  
\author{Denis Ullmo}
\affiliation{Univ.\ Paris-Sud, LPTMS UMR 8626, 91405 Orsay Cedex, France}
\affiliation{CNRS, LPTMS UMR 8626, 91405 Orsay Cedex, France}    
\date{\today}

\begin{abstract}
We provide a semiclassical description of the electronic transport through graphene $n$-$p$ junctions in the quantum Hall regime. This framework is known to experimentally exhibit conductance plateaus whose origin is still not fully understood. In the magnetic regime ($E < v_F B$), we show the conductance of excited states is essentially zero, while that of the ground state depends on the boundary conditions considered at the edge of the sample. In the electric
regime ($E > v_F B$), for a step-like electrostatic potential (abrupt on the scale of the magnetic length), we derive a semiclassical approximation for the conductance in terms of the various snake-like trajectories at the interface of the junction. For a symmetric configuration, the general result can be recovered using a simple scattering approach, providing a transparent analysis of the problem under study. We thoroughly discuss the semiclassical predicted
behavior for the conductance and conclude that any approach using fully phase-coherent electrons will hardly account for the experimentally observed plateaus.
\end{abstract}

\pacs{73.22.Pr, 73.43.Jn, 03.65.Sq, 73.23.Ad}

\maketitle

\section{Introduction}

Graphene, a two-dimensional material made of a mono-layer of carbon atoms arranged in a hexagonal lattice, has been studied theoretically for some time \cite{Wallace47,McClure56,Slonczewski58,Semenoff84}, but means to isolate and manipulate it experimentally have only been developed a few years ago \cite{Novoselov04}. Since then, the exotic properties of this material have aroused considerable interest in the condensed matter physics, chemistry, and material science communities \cite{Geim09,CastroNeto09}. These properties mostly originate from graphene's peculiar low-energy band structure, both gapless and linear. Conduction and valence bands touch each other at the corners of the (hexagonal) first Brillouin zone, two of which ($K$,$K'$) are inequivalent. At low energies, the dispersion relation vanishes at these Dirac points and the effective Hamiltonian for electrons in graphene is that of a massless
pseudo-relativistic particle travelling at an effective ``speed of light" $v_F$. Here, the internal degree of freedom associated with graphene's two inequivalent sublattices plays the role of the spin (the true spin providing merely a degeneracy factor as long as spin-orbit interaction or magnetic impurities are neglected, as we shall do here).


Two of the most prominent features that distinguish graphene from conventional two-dimensional electron gases (2DEG) are the occurrence of Klein tunneling \cite{Katsnelson06,Cheianov06,Stander09,Young09}, i.e. the absence of backscattering of charge carriers when approaching potential barriers under normal incidence, and that of an anomalous quantum Hall effect (QHE) \cite{Novoselov05,Zhang05} stemming from the peculiar quantization of the Landau levels in this material. 

A nice experiment probing the physics that emerges from the interplay of Klein tunneling and the anomalous QHE was carried out in 2007 by Williams and collaborators \cite{Williams07}. Using a global back gate and a local top gate, they were able to apply gate voltages of different signs in two regions of a graphene sheet, creating a $n$-$p$ junction at the regions interface, an ideal setup to study Klein tunneling. Using a strong perpendicular magnetic field they performed quantum Hall transport measurements across this graphene $n$-$p$ junction. Quite remarkably, for some combinations of $n$ and $p$ filling factors, the experimental data exhibit conductance plateaus at the fractions $1/2$ and $3/4$ of the quantum of conductance of spin-degenerate systems, $G_0 = 2 e^2/h$, at odds with the usual graphene quantum Hall series.

These observations were confirmed by other experiments \cite{Lohmann09,Ki09,Ahlers11} and extended to the setup of single-layer $n$-$p$-$n$ junctions \cite{Ozyilmaz07} and bilayer $n$-$p$-$n$ junctions \cite{Jing10}. The observed plateaus in $n$-$p$ junctions can be cast in terms of the QHE edge channel picture, within good accuracy,  by the expression  
\begin{equation}
\label{RMTfmm}
G_{np}  = G_0 \frac{N_n N_p}{N_n + N_p} \; ,
\end{equation}
where $N_n$ and $N_p$ stand for the number of incoming/outgoing edge modes, which depend on the gate voltages applied to the $n$ and $p$ regions, and consequently on their filling factors.  

Very early, Abanin and Levitov \cite{Abanin07sci} proposed an interpretation for Eq.\ \eqref{RMTfmm} in terms of a ``quantum chaos hypothesis". The key element is that since $n$ and $p$ edge channels in the quantum Hall regime have opposite chirality (direction of propagation), they interfere with each other at the junction interface. The ``quantum chaos hypothesis'' amounts to assume that this interference effect gives rise to a mode-mixing mechanism that is
sufficiently strong that an incoming channel at the interface will have equally likely probabilities of leaving the interface in any of the outgoing channels, leading to a current partition consistent with Eq.\ \eqref{RMTfmm}. This picture is further motivated by the observation that the values of the conductance plateaus $G_{np}$ coincide with the average conductance $\langle G \rangle$, predicted by the random matrix theory (RMT) for chaotic systems
\cite{Baranger94,Jalabert94}.

Albeit very appealing, a complete interpretation of the experimental results by means of the quantum chaos hypothesis is still problematic. As already pointed out in Ref.~[\onlinecite{Abanin07sci}], RMT gives an average $\langle G \rangle = G_{np}$, but also predicts universal conductance fluctuations (UCF) of the order of $e^2/h$. Those are definitely not currently observed by experiments. UCF are ubiquitous in both quantum chaotic \cite{Baranger94,Jalabert94} and diffusive systems \cite{Altshuler85,Lee85}. They have been studied experimentally as well as theoretically for ordinary graphene flakes (see, for instance, Ref.~[\onlinecite{Mucciolo10}] for a review). In the quantum Hall regime, numerical investigations of graphene $n$-$p$ junctions \cite{Li08,Long08} and of graphene-superconductor junctions \cite{Sun09} showed that modifications of the disorder strength may cause small changes in the magnitude of the UCF but do not imply qualitative modifications of the transmission. These observations thus support the conclusion that in the absence of decoherence, the role of which still has to be established in the problem we are investigating, disorder by itself cannot explain the conductance plateaus observed experimentally in quantum Hall $n$-$p$ junctions.


Further insight is gained from another recent experimental work investigating the fingerprints of Klein tunneling in a graphene $p$-$n$-$p$ junction, this time in the presence of a weak magnetic field \cite{Young09}. In this case, the authors observe beautiful quantum interference patterns {\it quantitatively} explained by a semiclassical approximation in terms of a set of simple transmitted and reflected paths across the junction. It is noteworthy that,
although the sample quality of all cited experiments is similar, in Ref.~[\onlinecite{Young09}] there is no indication that disorder plays a significant role. Obviously, one should be cautious before drawing conclusions from this observation, since by going from small to large magnetic fields one dramatically changes the physics of the problem. Nonetheless, it certainly justifies the study of the transmission properties of clean $n$-$p$ junctions in the quantum Hall regime.  

An even stronger motivation to consider a model of clean ballistic graphene is that, at the {\it classical} level, the snake-like\,\footnote{The expression ``snake" is borrowed from the works carried in the framework of variable magnetic fields \cite{Muller92,Rakyta08}.} trajectories of electrons on a graphene step-like symmetric $n$-$p$ junction have equal reflection and transmission probabilities (see
beginning of section \ref{secElecAmb}), leading to current partition consistent with the experiments.

The goal of this paper is to use a semiclassical analysis to investigate whether the classical current partition behavior is transposed into the ``quantum chaos hypothesis" in the quantum coherent transport limit. Extending over previous results already presented in \cite{Carmier10a}, we shall see that the answer is actually negative, and that clean graphene $n$-$p$ junctions in a high magnetic field provide an example where the classical behavior is not a good approximation to the semiclassical (or quantum) behavior. We find that, in the full quantum coherent transport regime, the transmission fluctuations due to quantum interference between different electronic paths across the junction are large and non-universal. The relation between these results and the experimental ones is discussed in sections \ref{secDisc} and \ref{secConc}.	

Our analysis is based on the Landauer conductance formula
\begin{equation}
\label{LanButt}
G = G_0 \sum_{n,\alpha} T_{n,\alpha},
\end{equation} 
where $T_{n,\alpha}$ is the transmission coefficient, $n$ stands for the channel index, and $\alpha$ for the valley index that specifies in which of the valleys $K \; (\alpha = +1)$ or $K' \; (\alpha = -1)$ the edge channel is polarized\,\footnote{In fact, polarization of the edge states in a given valley occurs for infinite mass and zigzag boundary conditions, but not for the armchair case where edge channels are polarized in a linear superposition of valleys \cite{Brey06qhe,Abanin07ssc}, \cite{Beenakker08}.}. We evaluate the transmission coefficients via the Green's functions formalism introduced by Fisher and Lee \cite{Fisher81} and later generalized by Baranger and Stone \cite{Baranger89} (mainly to include the presence of a magnetic field). Here we use semiclassical Green's functions obtained for graphene by two of us in a previous publication \cite{Carmier08}. The semiclassical approximation allows us to express $G$ analytically and can be used to model systems of realistic sizes, which are hardly reachable by numerical techniques that use an atomistic basis.

The description of a graphene $n$-$p$ junction depends sensitively on the relative strength of the magnetic and electric fields at the interface between the electron-doped and ``hole"-doped regions. This is best understood by recalling the insight gained in graphene electronic transport in presence of crossed electric (in plane) and magnetic (perpendicular) fields \cite{Shytov07,Shytov09}: the electromagnetic Lorentz invariant quantities ${\cal L}_1 = {\bf E} \cdot {\bf B}$ and ${\cal L}_2 = {\bf E}^2 - v_F^2 {\bf B}^2$ define two distinct transport regimes characterized by the parameter
\begin{equation} \label{eq:beta}
\beta = \frac{v_F|{\bf B}|}{|{\bf E}|} \; .
\end{equation}
If $\beta < 1$ the electric field will always dominate the magnetic field, as it is possible to apply a Lorentz boost and go to a reference frame where the magnetic field is made zero. This is referred to as the electric regime. In contrast, if $\beta > 1$ the electric field can be eliminated by going to a proper reference frame. This regime is referred to as the magnetic one. The Lorentz invariance stated above relies on translational invariance, and hence is not expected to hold near the edges of a graphene ribbon. Nonetheless, the classification in terms of electric and magnetic regimes still makes sense for confined systems, as will be shown in the following.

The remainder of the paper is organized as follows. We begin presenting our model and the key elements of the semiclassical analysis. In section \ref{secMag}, we then discuss the physics of the magnetic regime (in particular the limit $\beta \gg 1$). Within this regime (for which no quantum chaos hypothesis has been evoked), we show that a classical description based on the concept of adiabatic invariance suffices to explain most of the junction's features. We obtain that the conductance is essentially zero for all modes except for the lowest one, whose transmission depends on the boundary conditions considered at the edge of the ribbon \cite{Tworzydlo07,Akhmerov08vv}. 

The electric regime, which we believe to be that of the experiments \cite{Williams07,Lohmann09,Ki09,Ozyilmaz07}, is addressed in sections \ref{secElecAmb} and \ref{secElecGen}. Those constitute the main technical development of this paper. At their end, we present a summary and a discussion of the main results they contain. In section \ref{secElecAmb} the electric potential is taken to be symmetric, such that the (absolute) doping is the same at the $n$ and $p$ regions. This
allows for a simple and transparent analysis using a semiclassical scattering matrix approach. A general treatment, beyond the symmetric case, of the electric regime is presented in section \ref{secElecGen}. It requires determining the exhaustive list of trajectories involved in the dynamics at the interface, and making use of the semiclassical approximation to the single particle Green's function in graphene \cite{Carmier08} to compute the Fisher-Lee formulae. 
We present generalized expressions of the latter, which take into account the internal pseudo-spin structure of charge carriers in graphene. Some qualitative features of the transmission in the general case are similar to the ones found in the simpler symmetric configuration. In particular, in both cases the conductance differs significantly from the classical one. A qualitative analysis of the robustness of our semiclassical findings with respect to modifications of the model such as boundary conditions, disorder and steepness of the potential barrier is conducted in section \ref{secDisc}. Our conclusions are presented in section \ref{secConc}.

\section{The Model}
\label{secModel}

We consider a graphene sheet of finite width $W$ in the $y$ direction, and connected in the $x$ direction to infinite leads on both sides. We assume the $n$-$p$ junction is obtained by applying an electrostatic potential in the plane of the graphene sheet, separating it in three distinct regions:
\begin{equation} \label{eq:regs}
V(x) =\left\{
\begin{array}{l}
V_1 \; , \; x < 0 \; (\text{region 1})
\vspace*{0.3cm}
\\
V_1 + eEx \; , \; 0 < x < L \; (\mbox{region } \mathcal{J})
\vspace*{0.3cm}
\\
V_2 = V_1 + \Delta \; , \; x > L \; (\text{region 2})
\end{array}
\right. \; ,
\end{equation}
with $V_1 < 0$, $V_2 > 0$ (we assume the chemical potential fixed at 0), and $\Delta = eEL$ the height of the potential step. $L$ governs the steepness of the potential step and $E$ is the intensity of the associated electric field ${\bf E} = E\hat{{\bf x}}$, where $\hat{{\bf x}}$ is the unit vector in the $x$ direction. Applying a strong perpendicular magnetic field ${\bf B} = B\hat{{\bf z}}$, such that the quantum Hall regime $l_B \ll W$ ($l_B=\sqrt{\hbar/(eB)}$ is the magnetic length) is reached, leads to the classical picture of skipping-orbit motion at the edges of regions 1 and 2, with a direction of propagation depending on the edge and the type of charge carriers. The electron dynamics in region $\mathcal{J}$ and in its neighborhood is what interests us here.

To describe the quantum dynamics of the electrons in the graphene sheet, we use the representation ${\bf \Psi} = (\psi_A,\psi_B,-\psi'_B,\psi'_A)^T$ for the wave function. Here, as usual, $(A,B)$ denote the real-space graphene sublattices, $(\psi_A,\psi_B)^T$ the wave-function in valley $K$ and $(\psi'_A,\psi'_B)^T$ that in valley $K'$. In such a representation the low-energy Hamiltonian of graphene electrons is isotropic in valley space and takes the form
\begin{equation}
\label{Ham44gra}
{\cal H} = v_F\tau_0\otimes(\hat{{\bf \Pi}}\cdot\vec{\sigma}) +
 U({\bf r})\tau_0\otimes\sigma_0 + m({\bf r}) v_F^2 \tau_z\otimes\sigma_z \; ,
\end{equation}
where $U({\bf r})$ is the electric potential generating $\bf E$, the magnetic field $\bf B$ derives from the vector potential ${\bf A}({\bf r})$ via the Peierls substitution $\hat{{\bf \Pi}} = \hat{{\bf p}} + e{\bf A}({\bf r})$, and a mass term $m({\bf r})$ has been included for completeness. In Eq.~(\ref{Ham44gra}) $\vec{\tau}$ and $\vec{\sigma}$ are Pauli matrices respectively associated to valley space and sublattice space (with the usual convention that $\tau_0 = \sigma_0 = 1\!\!1$), and the scalar product is restricted to the $x$-$y$ graphene plane.

Three types of boundary conditions will be discussed: most of the time we shall assume infinite mass confinement, but also in some circumstances zigzag and armchair. All of these can be incorporated in a matrix representation introduced by Akhmerov and Beenakker \cite{Akhmerov08,Beenakker08}, 
\begin{equation}
\label{boundAB}
\begin{array}{l}
{\bf \Psi} = {\cal M}{\bf \Psi} \; ,
\vspace*{0.2cm}
\\
{\cal M} = (\vec{\nu}\cdot\vec{\tau})\otimes(\vec{n}\cdot\vec{\sigma}) \; ,
\end{array}
\end{equation}
where $\vec{n}$ is the normal to the outward pointing unit vector at the boundary and $\vec{\nu}$ is the polarization of the edge state in valley space. Each boundary condition is associated to a given matrix ${\cal M}$, which actually amounts to a given polarization $\vec{\nu}$. For instance, the case of infinite mass confinement corresponds to imposing the condition $\Psi_B/\Psi_A = \Psi'_B/\Psi'_A= i\exp(i\gamma)$ on the boundary with $\gamma=\cos^{-1}(\vec{n}\cdot \hat{{\bf y}})$, as shown by Berry and Mondragon \cite{Berry87}. This is obtained by choosing $\vec{\nu} = \hat{{\bf z}}$ in Eq.~(\ref{boundAB}). Similar parameterizations are obtained for zigzag and armchair lattice terminations (see Refs.~\cite{Akhmerov08,Beenakker08}).

\begin{figure}
\begin{center}
\includegraphics[width=0.8\linewidth]{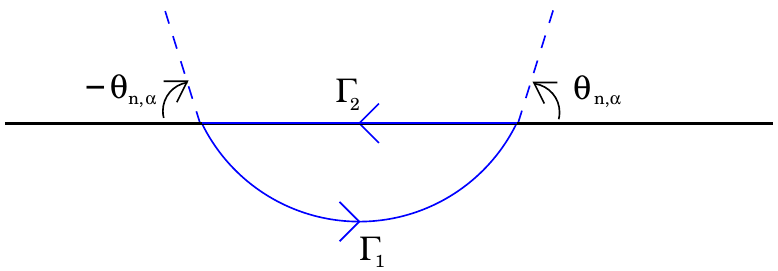}
\caption{(Color online) Illustration of a closed path $\Gamma = \Gamma_1\cup\Gamma_2$
  along which the action integral in Eq.~(\ref{EBKgra}) is
  computed. The phase of the mode wave-function must remain unchanged
  (up to an integer multiple of $2\pi$), leading to the quantization
  of the edge angle $\theta_{n,\alpha}$.}  
\label{FigTrajferm}
\end{center}
\end{figure}

At the leads, that is, within regions 1 and 2 of Eq.~\eqref{eq:regs}, a propagating mode $(n,\alpha)$ in the quantum Hall regime can be built on the set of skipping trajectories emerging from the boundary with an angle $\theta_{n,\alpha}$ such that the transverse action is quantized, according to the generalized EBK formula \cite{Keppeler02}
\begin{equation}
\label{EBKgra}
\frac{1}{\hbar}\oint {\bf p}\cdot d{\bf r} - \mu\frac{\pi}{2} + \xi_{sc} + \phi_{bc} = 2\pi n \quad .
\end{equation}
In this quantization condition, the closed path on which the integration is performed can be taken as the one shown on Fig.~\ref{FigTrajferm} (bearing in mind that on $\Gamma_2$ the path does not follow a trajectory, and thus the momentum is not parallel to the shown direction), $\mu$ is a Maslov index counting the caustics traversed by the closed path and $\phi_{bc}$ is the phase acquired by the electron upon bouncing on the boundary. For graphene, one needs in addition to take into account a semiclassical phase $\xi_{sc}$\ which, in the absence of a mass term in the Hamiltonian (\ref{Ham44gra}) coincides with a Berry phase \cite{Carmier08}, and is just given by half the rotation angle of the momentum vector (see the discussion at the beginning of section \ref{secElecGen}). Because inside the graphene ribbon we only consider a massless Hamiltonian, in what follows we refer to the semiclassical phase as a Berry phase, for simplicity.

For infinite mass boundary conditions, the phase acquired upon reflection on the boundary $\phi_{bc} = \phi_{m_\infty}$ in Eq.~(\ref{EBKgra}) can be easily shown to be $\phi_{m_\infty} = \pi$ (Dirichlet) in valley $K$ and $\phi_{m_\infty} = 0$ (Neumann) in valley $K'$. The path of integration $\Gamma$ traverses one caustic (so that $\mu=1$) and $\xi_{sc}^{\Gamma_1} = -\xi_{sc}^{\Gamma_2}$. Wrapping everything up, the quantization condition can be written as
\begin{equation}
\label{quantS1}
eBR_i^2 \left( \theta_{n,\alpha} - \frac{\sin{2\theta_{n,\alpha}}}{2}
\right) = 2\pi\hbar(n - \frac{\alpha}{4}) 
\end{equation} 
with 
\begin{equation}
\label{eq:R}
R_i = \frac{|V_i|}{eBv_F}
\end{equation} 
the cyclotron radius at the electron ($i=1$) or hole ($i=2$) side. Note the quantization condition (\ref{quantS1}) with $n = 0$ can be fulfilled for only one of the two valleys ($\alpha=-1$, for which Neumann boundary conditions apply). Introducing the function
\begin{equation}
\label{functionf}
f : \theta \to \theta - \sin{(2\theta)}/2 \; ,
\end{equation}
and noting that $e B R_i^2/\hbar = 2 \nu_i$, where 
\begin{equation} \label{eq:filling}
   \nu_i = \frac{(k^{(i)}_F l_B)^2}{2} 
\end{equation}
is the filling factor at the electron ($i\!=\!1$) or hole ($i\!=\!2$) region, we  recast Eq.~(\ref{quantS1})  as 
\begin{equation}
\label{quantS2}
\frac{1}{\pi} f(\theta_{n,\alpha}) = (n - \frac{\alpha}{4})\nu_i^{-1} \; .
\end{equation} 
This form of the quantization condition makes explicit that, besides the indices $(n,\alpha)$, the quantized angle $\theta_{n,\alpha}$ only depends on the filling factor $\nu_i$. Quantization conditions for zigzag and armchair boundary conditions can be derived in the same way \cite{Rakyta09}. Note finally that as the angle $\theta_{n,\alpha} \to \pi$, which corresponds to a transition to a Landau level in the bulk (for which $\alpha$ can be interpreted as switching to $0$) a uniform approximation should be used. This uniform approximation is described for the scalar (Schr\"odinger) case in Ref.\ \cite{Avishai08}.

\section{Magnetic regime}
\label{secMag}

In this section, we study the transport in the magnetic regime in the limit of $\beta \gg 1$. 
\begin{figure}
\begin{center}
\includegraphics[width=0.99\linewidth]{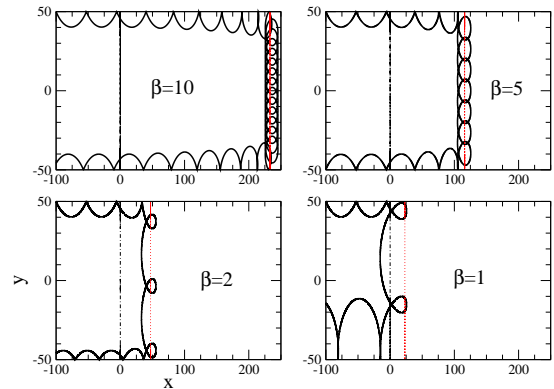}
\caption{(Color online) Skipping orbit motion for classical electrons coming from the 
  upper left edge with an angle $\theta=\pi/4$, followed by bubbling. Different panels 
  correspond to various values of $\beta$ in the magnetic regime. The black dot-dashed 
  lines indicate where the junction region begins and the red dotted lines indicate where 
  bubbling is expected to occur according to the semiclassical theory. The adiabatic 
  transition from the edge to the bulk is apparent in the upper panels, while it clearly 
  breaks down in the lower ones even though classical bubbling still occurs.} 
\label{FigBubbling}
\end{center}
\end{figure}
This corresponds to the situation where the electric potential varies adiabatically on the scale of the cyclotron radius. One can then use that the action (\ref{quantS1}) of the electron along the edge is an adiabatic invariant, and is thus conserved while propagating in the $x$ direction. Defining an effective local cyclotron radius $R(x) = |V(x)|/(eBv_F)$, conservation of the action implies that the decrease in $R(x)$ must be compensated by an increase in the effective angle $\theta(x)$ along the edge, since the function $f(\theta)$ in Eq.~(\ref{functionf}) is monotonous on the interval [$0$,$\pi$]. When this angle reaches the value $\pi$, a transition takes place from the edge skipping-orbit motion to the transverse bulk cyclotronic motion with a vertical drift caused by the electric field. We will henceforth refer to this transition by the visually appropriate term of ``bubbling". Once an electron has bubbled from the edge, it traverses the graphene ribbon transversally until it reaches the opposite edge, and then propagates back into the lead that it came from. This scenario is illustrated in Fig.~\ref{FigBubbling}.

In the case of a $n$-$n$ junction, we see that the transmission probability of an electron can only be $0$ or $1$. Quantitatively, the bubbling abscissa of an electron is determined by the relation $R_1^2 f(\theta_{n,\alpha}) = R^2(x_{bub})\pi$, which leads to the value $x_{bub} = \beta R_1 (1 - \sqrt{f(\theta_{n,\alpha})/\pi})$. A given mode $(n,\alpha)$ bubbles if $x_{bub} < L$, which can be equivalently expressed through the bubbling condition
\begin{equation}
\label{BubCond}
-\frac{V_2}{|V_1|} < \sqrt{\frac{f(\theta_{n,\alpha})}{\pi}} 
\end{equation}
(which of course is always realized for $n$-$p$ junctions). 
\begin{figure}
\begin{center}
\includegraphics[width=1.\linewidth]{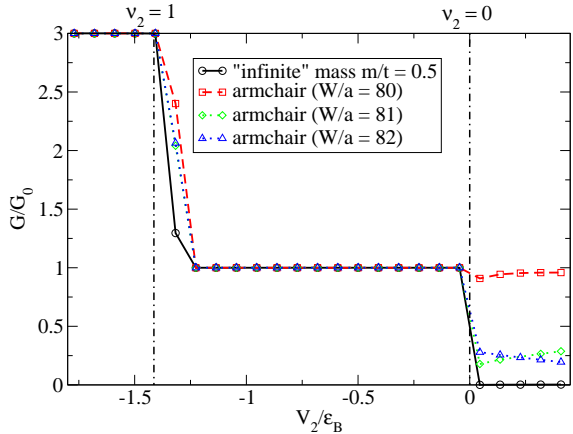}
\caption{(Color online) Exact (quantum mechanical) conductance $G$ across the junction 
  as a function of the potential $V_2$ (in units of $\epsilon_B = \hbar v_F/l_B$) in the 
  right side of the junction, with a fixed value $-V_1/\epsilon_B = 1.82$ of the potential 
  in the left side of the junction (which in terms of the filling factor is $\nu_1=1.65$). 
  Solid line (black online): ``infinite'' mass boundary condition (note the mass actually 
  used in the numerical simulation is $m=0.5 t$, with $t$ the hopping parameter); Dashed 
  (red online): armchair with $W/a=80$ ($a$ is the distance between nearest carbon atoms 
  {\em of the same sublattice}); Dotted with rhomb symbols (green online): armchair with 
  $W/a=81$; Dotted with triangle symbols (blue online): armchair with $W/a=82$. The vertical 
  dot-dashed line indicates values of $V_2$ for which the filling ratio $\nu_2$ in region 2 
  is an integer. When $V_2 < 0$ the system is effectively a $n$-$n$ junction, and the 
  conductance is given by the number of channels in region 2, namely $2[\nu_2]+1$ since 
  $\text{min}([\nu_1],[\nu_2])=[\nu_2]$ here (see discussion in main text). For positive 
  $V_2$, the conductance across the $n$-$p$ junction is governed by the transmission of the 
  lowest channel and, as expected according to Eq.~(\ref{Mono}), is given by $G = G_0$ for 
  armchair boundary conditions with $W/a = 2\;(\text{modulo}\;3)$, $G = G_0/4$ for other 
  armchair widths and $G=0$ for the infinite mass confinement case. The plot was obtained 
  for $\beta=5$, with a system of length $L/a = 60$ and a magnetic length $l_B/a = 6.8$.} 
\label{Fig:Magnetic}
\end{center}
\end{figure}
\begin{figure}
\begin{center}
\includegraphics[width=1.\linewidth]{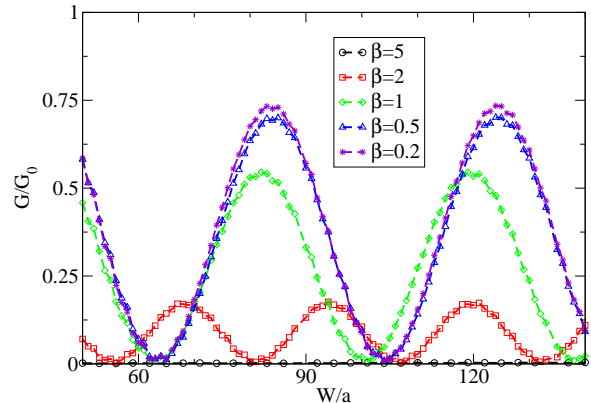}
\caption{(Color online) Exact (quantum mechanical) conductance $G$ across the junction 
  as a function of the width $W$ (in units of $a$) for different values of the parameter 
  $\beta$ governing the transition between the magnetic ($\beta \gg 1$) and the electric 
  ($\beta \ll 1$) regimes. The system considered was confined by an ``infinite'' mass 
  $m/t = 0.5$ and the filling factors on both sides of the junction were taken to be 
  $\nu_1=1.65$ and $\nu_2=0.10$ (hole doped). The length of the system and the value of 
  the magnetic field are the same as in the previous figure. We observe that the adiabatic 
  result $G \simeq 0$ expected from Eq.~(\ref{Mono}) is observed for $\beta=5$, but that 
  $\beta=2$ already shows some oscillations in the conductance. For $\beta\leq1$, the 
  featured oscillations resemble those predicted by our semiclassical calculations in 
  sections \ref{secElecAmb} and \ref{secElecGen}.}
\label{Fig:MassBeta}
\end{center}
\end{figure}
This leads naturally for a $n$-$n$ junction to the result that $G = G_0 \text{min}(N_1,N_2)$ \cite{Abanin07sci}, with $N_i = 2[\nu_i]+1$ (with $[\cdot]$ the integer part), as expected in the context of quantum adiabatic transport \cite{Beenakker91}. Note that the quantum adiabaticity criterion, namely that the electrostatic potential must vary less than the inter-Landau level spacing on the scale of the magnetic length, is actually more robust that the classical one.

For a $n$-$p$ junction, transmission is zero for all modes, except possibly for the zero-energy mode for which the semiclassical reasoning above cannot be applied. The latter must be treated separately. Hence, a $n$-$p$ junction in the adiabatic limit $\beta \gg 1$ has a conductance $G=G_0 T_0$, with $T_0$ the transmission probability of the $n = 0$ mode.

In the absence of inter-valley scattering, the transmission coefficient $T_0$ of a single channel through a $n$-$p$ junction in the quantum Hall regime depends only on the valley-polarization of the edge states \cite{Tworzydlo07,Akhmerov08vv}, namely
\begin{equation}
\label{Mono}
T_0 = \frac{1 - \cos{\phi_\nu}}{2},
\end{equation}
where the angle $\phi_\nu$ is the one separating the valley-polarization vectors of top and bottom edge states on the Bloch sphere. For an armchair ribbon, this leads to plateaus in the conductance at $G_0$ or $G_0/4$ depending on the width of the ribbon \cite{Tworzydlo07}. For a zigzag ribbon, it was realized that formula (\ref{Mono}) cannot be applied since a potential barrier, no matter how smooth it is, causes inter-valley scattering \cite{Akhmerov08vv}. However, very similarly to the armchair case, it turns out that the transmission depends on the width of the ribbon and can be either zero or one\,\footnote{These plateaus have not yet been observed experimentally, probably due to the existence of valley-mixing edge roughness in the samples \cite{Low09b}.}. The case of infinite mass confinement can be treated similarly to the armchair one, yielding zero transmission just like for the higher modes. These results for the armchair as well as infinite mass confinement cases are illustrated on Fig.~\ref{Fig:Magnetic}. The perfect reflection for infinite mass confinement is a trivial consequence of the conservation of the valley-polarization of the edge state. The expected result (in the limit $\beta \gg 1$) for the conductance of a graphene $n$-$p$ junction laterally confined by an infinite mass is thus zero, up to exponentially small tunneling contributions. As illustrated in Fig.~\ref{Fig:MassBeta}, this limit is already reached from this point of view for $\beta \simeq 5$, when a $\beta \simeq 2$ is already in the transition towards the electric regime where some oscillations in transmission (and thus the conductance) are already visible. The data presented in Figs.~\ref{Fig:Magnetic},\,\ref{Fig:MassBeta} were obtained with a recursive Green's function technique, using the numerical software KNIT developed by Kazymyrenko and Waintal \cite{Kazymyrenko08}.

\section{Electric regime: symmetric case}
\label{secElecAmb}

We switch now to  the electric regime. Two important simplifications are made in the next couple of sections. 
\begin{figure}
\begin{center}
\includegraphics[width=0.6\linewidth]{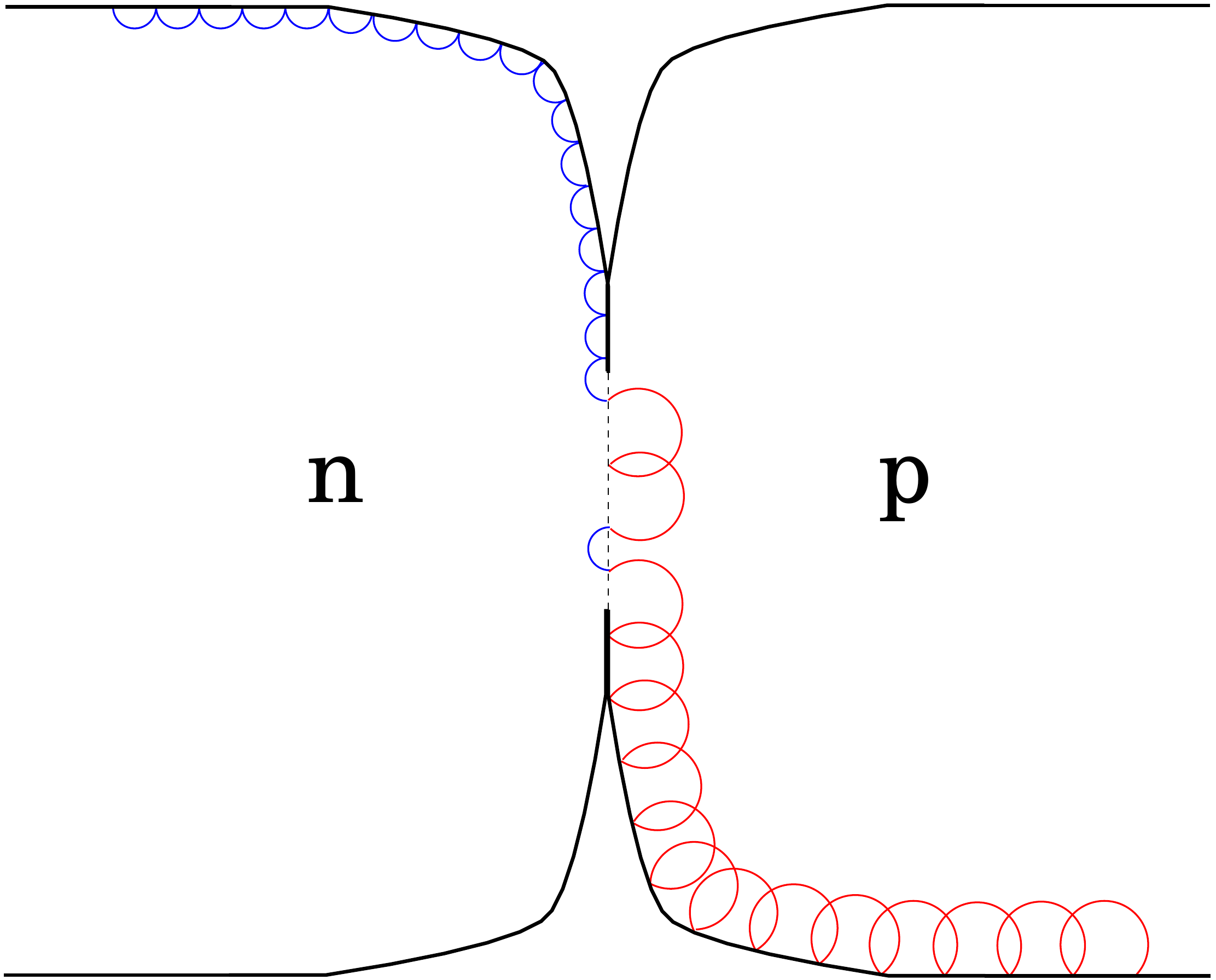}
\caption{(Color online) Sketch of the geometry of a parallel junction. Due to the 
  smooth transition from the edge of the ribbon to the $n$-$p$ step-like junction, 
  the quantization of the incoming channels is maintained.}
\label{FigParAdia}
\end{center}
\end{figure}
First, we consider the electric potential to be step-like on the scale of the magnetic length (but not on that of the carbon lattice, so as to avoid inter-valley scattering), placing ourselves in the opposite limit ($\beta \ll 1$) as that of section~\ref{secMag}. This makes it possible to neglect the magnetic field during the interaction with the barrier. The second simplifying hypothesis has to do with geometry. Compared to the experimental setup, we will consider a graphene ribbon where the transition from the edge to the step-like junction is very smooth (as in Fig.~\ref{FigParAdia}), such that quantization of the edge modes is maintained when arriving on the $n$-$p$ interface. This amounts to taking the junction ``parallel'' to the edge of the ribbon instead of perpendicular to it. The main reason for making this choice is of course that the parallel junction is a simpler problem to tackle analytically. However, as discussed in more details in \cite{Carmier11c}, it can be shown that except for diffractive-like contributions at the edge-junction corner, going from a geometry for which the junction is perpendicular to the edges of the ribbon to one where it is parallel mainly amounts to applying a unitary transformation to the mode basis, under which the Landauer-B\"uttiker formula (\ref{LanButt}) is invariant. As in any case a completely realistic description of the dynamics in the corners would depend on many details not included in our model (and probably unknown), the parallel junction is presumably as close (or as far) from a perfectly realistic description of the junction than a perpendicular one.

\begin{figure}
\begin{center}
\includegraphics[width=0.99\linewidth]{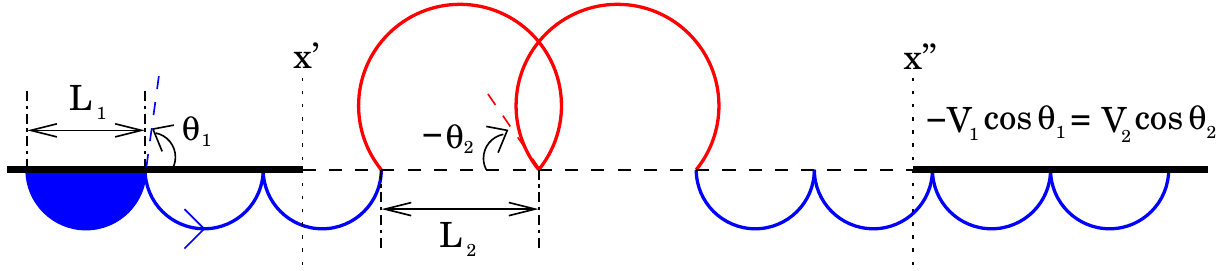}
\caption{(Color online) Typical trajectory along a parallel junction (rotated by 
  an angle $\pi/2$ with respect to Fig.~\ref{FigParAdia}) between initial $x = x'$ 
  and final $x = x''$ Poincar\'e sections. Excursions cover a length $L_1$ on the 
  electron side and $L_2$ on the hole side. Angles with respect to the interface 
  axis are shown.}
\label{FigBetPoin}
\end{center}
\end{figure}

We start by introducing a few notations for trajectories such as the one illustrated on Fig.~\ref{FigBetPoin}. Let us denote by $\theta_1$ and $\theta_2$ the angles between the $x$-axis and the vector $\hat{{\bf \Pi}} = \hat{{\bf p}} + e{\bf A}({\bf r})$ (which is parallel to the velocity in the $n$ region, but {\em antiparallel} to it in the $p$ region) when the trajectory emerges from the edge or from the junction in respectively the $n$ and $p$ side. The two
angles are related through the Snell-Descartes law
\begin{equation}
\label{conserv}
|V_1|\cos{\theta_1} = V_2\cos{\theta_2}
\end{equation}
which expresses conservation of the momentum in the $x$ direction. During the time interval between two consecutive bounces on the edge boundary or at the junction interface (we call this an ``excursion"), an electron covers a distance $L_1 = 2R_1\sin{\theta_1}$ on the $n$ side (respectively $L_2 = 2R_2\sin{\theta_2}$ on the $p$ side) in the longitudinal $x$ direction.

Klein tunneling or reflection at the junction interface is given by the same probability amplitude as in the absence of magnetic field, namely \cite{Katsnelson06}
\begin{eqnarray}
   \tilde r_1  & = & e^{i\theta_1} r_1
   \qquad \qquad \mbox{for reflection $1 \to 1$} \label{eq:tr1} \\
   \tilde t_1 & = & e^{i(\theta_1 + \theta_2)/2} t_1
   \qquad \hskip-0.27cm \mbox{for transmission $1 \to 2$} \; , \label{eq:tt1}
\end{eqnarray}
with 
\begin{eqnarray}
   r_1  & = & - \frac{\cos((\theta_1 + \theta_2)/2)
   }{\cos((\theta_1 - \theta_2)/2)}  \label{eq:r1} \\
   t_1 & = & -i   \frac{\sin{\theta_1}}{\cos{((\theta_1 - \theta_2)/2)}} 
   \label{eq:t1} \; .
\end{eqnarray}
For incident waves from the $p$ side, the corresponding expressions for $(\tilde r_2,\tilde t_2)$ and $(r_2,t_2)$ are obtained by exchanging the roles played by $\theta_1$ and $\theta_2$. The phases of the factors $e^{i\theta_1}$ and $e^{i(\theta_1 + \theta_2)/2}$ in (\ref{eq:tr1}) and (\ref{eq:tt1}) can be interpreted as Berry phases as they correspond to half the pseudo-momentum (which as already mentioned is, for holes, {\em antiparallel} to that of the velocity) rotation during the scattering on the junction interface ($\theta_1 = -(-\theta_1 - \theta_1)/2$ and $(\theta_1 + \theta_2)/2 = -(-\theta_2 - \theta_1)/2$). We have therefore distinguished them from the ``genuine'' reflection and transmission coefficients $r_1$ and $t_1$ given by (\ref{eq:r1}) and (\ref{eq:t1}).

The transmission probability through the interface is deduced from the quantum amplitude by taking into account the flux normal to the barrier:
\begin{equation}
T = \frac{\sin{\theta_2}}{\sin{\theta_1}}|t_1|^2 
= \frac{\sin{\theta_1}}{\sin{\theta_2}}|t_2|^2 = 
\frac{\sin{\theta_1}\sin{\theta_2}}{\cos^2{((\theta_1 - \theta_2)/2)}} \; .
\end{equation}

\begin{figure}
\begin{center}
\includegraphics[width=0.99\linewidth]{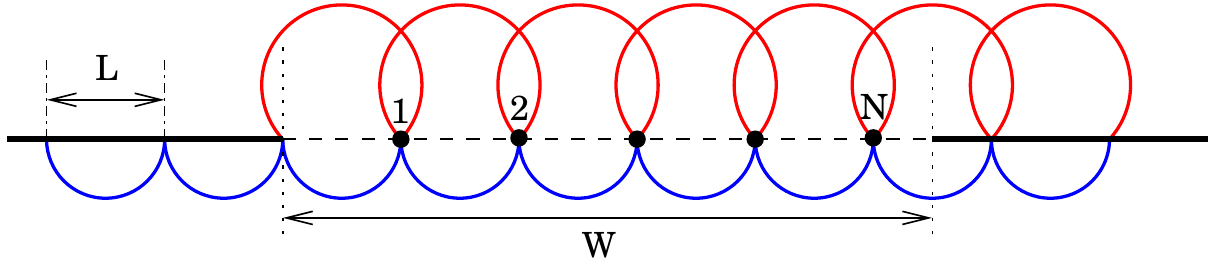}
\caption{(Color online) In the symmetric case, the length $L$ of excursions on 
  both sides of the junction are equal, making reflected and transmitted waves 
  meet at equidistant vertices along the interface. $N$ is the largest number 
  of excursions for a given interface length $W$.}
\label{FigAmbiSout}
\end{center}
\end{figure}

The rest of this section will be devoted to the symmetric case $V_2 = -V_1$. This leads via Eq.~(\ref{eq:R}) to $R_1=R_2$ and via Eq.~(\ref{conserv}) to $\theta_2 = \theta_1$, and hence $L_2 = L_1$. Excursions on both sides of the interface cover the same distance, implying that the reflected and transmitted waves of a scattering charge carrier meet at equidistant ``vertices" (see Fig.~\ref{FigAmbiSout}).

Let us start by quickly discussing what is expected classically for this configuration. A classical incident electron has a probability $T = \sin^2{\theta}$ of being transmitted through the (symmetric) potential step. Calling $X_i = (e_i,h_i)^T$ the vector composed of the probabilities $e_i$ and $h_i$ for the incoming particle to emerge at vertex $i$ on the $n$ or $p$ sides of the junction ($e_i + h_i = 1$) we have
\begin{equation}
\label{ScattClass}
X_{i + 1} = \left( \begin{array}{cc} 1 - T & T \\ T & 1 -
    T \end{array} \right) X_i \; .
\end{equation}
The matrix in Eq.~(\ref{ScattClass}) can be diagonalized and has eigenvalues $1$ and $\lambda = 1 - 2T$, which leads to  
\begin{equation}
\label{ClassN}
X_N = \frac{1}{2} \left( \begin{array}{cc} 1 + \lambda^N & 1 - \lambda^N \\ 
1 - \lambda^N & 1 + \lambda^N \end{array} \right) X_0 \; .
\end{equation}
Since $|\lambda| < 1$, the asymptotic behavior is as expected $X_\infty = (1/2,1/2)^T$ and Eq.~(\ref{ClassN}) tells us this limit is reached exponentially quickly with the number of bounces on the interface. An incoming classical electron has thus equal chances of being reflected or transmitted, provided the interface between regions $1$ and $2$ is long enough. Let us now show that this is no longer the case for a quantum particle.

Semiclassically, one needs now to propagate the amplitudes $z^{(e)}$ and $z^{(h)}$ on the electron and hole sides from one vertex to the other. Noting $Z_i = (z^{(e)}_i,z^{(h)}_i)^T$ these amplitudes at vertex $i$, this propagation can be obtained as $Z_{i + 1} = {\cal S}Z_i$, where the scattering matrix can be written as a product ${\cal S} = {\cal P} {\cal R}$ with
\begin{equation} {\cal P} = \left( \begin{array}{cc}
      e^{\frac{i}{\hbar}S_1 - i\frac{\pi}{2}\mu_1 + i\xi_1} & 0 \\ 0 &
      e^{\frac{i}{\hbar}S_2 - i\frac{\pi}{2}\mu_2 +
        i\xi_2} \end{array} \right)
\end{equation}
describing the propagation in the $n$ and $p$ regions and  
\begin{equation}
{\cal R} = \left( \begin{array}{cc} \tilde r_1 & \tilde t_2 \\ 
\tilde t_1 & \tilde r_2 \end{array} \right)
\end{equation}
the transmission or reflection taking place at the interface. The matrix ${\cal P}$ implies mainly a multiplication by a phase, which includes the action integral $S_i$ along the classical trajectory, the Maslov phase $\mu_i$ associated with the traversal of caustics, and the Berry phase $\xi_i$ associated with (half) the rotation of the pseudo-momentum vector $\Pi$. One obtains for these various quantities $S_1 = eBR_1^2 f(\theta_{n,\alpha})$, with the function $f$ defined in Eq.~(\ref{functionf}), $S_2 = S_1 - 2\pi\hbar\nu$ where $\nu=\nu_1=\nu_2$ is the filling factor (\ref{eq:filling}) in both the electron and hole regions, $\xi_1 = \xi_2 = -\theta_{n,\alpha}$ and $\mu_1 = - \mu_2 = 1$ (the Maslov index accounts for a single caustic and is counted negatively on the hole side since velocity and momentum are opposite there). For the symmetric junctions we consider here, we furthermore have $\tilde r_1 = \tilde r_2 = -e^{i\theta_{n,\alpha}}\cos{\theta_{n,\alpha}}$ and $\tilde t_1 = \tilde t_2 = -ie^{i\theta_{n,\alpha}}\sin{\theta_{n,\alpha}}$, so that finally
\begin{equation}
  \label{ScattSemi}
  {\cal S} = -e^{\frac{i}{\hbar}S_1 - i\pi \nu}
  \left( \begin{array}{cc}
      -ie^{i\pi\nu}\cos{\theta_{n,\alpha}} &
      e^{i\pi\nu}\sin{\theta_{n,\alpha}} \vspace*{0.2cm} \\
      -e^{-i\pi\nu}\sin{\theta_{n,\alpha}} &
      ie^{-i\pi\nu}\cos{\theta_{n,\alpha}} \end{array}
  \right) \; . 
\end{equation}
For a given channel ($n$,$\alpha$), the global phase factor in the scattering matrix expression is irrelevant and will henceforth be dropped. The resulting unitary matrix then has eigenvalues $\lambda_\pm = e^{\pm i \phi}$ with $\cos{\phi} = \cos{\theta_{n,\alpha}}\sin{(\pi \nu)}$.

Let us denote $N = [W/L]$ the integer part of the ratio of interface length $W$ and excursion length $L = L_1 = L_2$. Depending on its coordinate $y'$ in the initial Poincar\'e section $x = x'$, a charge carrier will bounce on the interface either $N$ or $N-1$ times. One can easily see the proportion of charge carriers bouncing $N$ times on the interface is given by the quantity $\{W/L\} = W/L - N$. Reflection and transmission probabilities for channel ($n$,$\alpha$) are then straightforwardly given by the simple expression
\begin{equation}
\label{ScattSol}
\left( \begin{array}{l} R_{n,\alpha} \\ T_{n,\alpha} \end{array}
\right) = \{W/L\} \left( \begin{array}{l} e_N \\
    h_N \end{array} \right) + (1 - \{W/L\})
\left( \begin{array}{l} e_{N - 1} \\ h_{N - 1} \end{array} \right) \; . 
\end{equation}
[This equation is valid actually both in the classical and semiclassical frameworks, but in this latter case with $e_i=|z^{(e)}_i|^2$, $h_i=|z^{(h)}_i|^2$]. From (\ref{ScattSemi}) we have
\begin{equation}
  e_N = \cos^2{(N\phi)} + C\sin^2{(N\phi)} \; , 
\end{equation}
\begin{equation}
h_N = \sin^2{(N\phi)} - C\sin^2{(N\phi)} \; ,
\end{equation}
with
\begin{equation}
C = \left(
  \frac{\sin^2{\theta_{n,\alpha}} -
    (\cos{\theta_{n,\alpha}}\cos{(\pi\nu)} -
    \sin{\phi})^2}{\sin^2{\theta_{n,\alpha}} +
    (\cos{\theta_{n,\alpha}}\cos{(\pi\nu)} - \sin{\phi})^2}
\right)^2 \; .
\end{equation}
Note that, for a fixed value of the indices $(n,\alpha)$, these quantities depend solely on the filling factor $\nu$ since, by the way, so does the angle $\theta_{n,\alpha}$ via Eq.~(\ref{quantS2}).

\begin{figure}
\begin{center}
\includegraphics[width=0.8\linewidth]{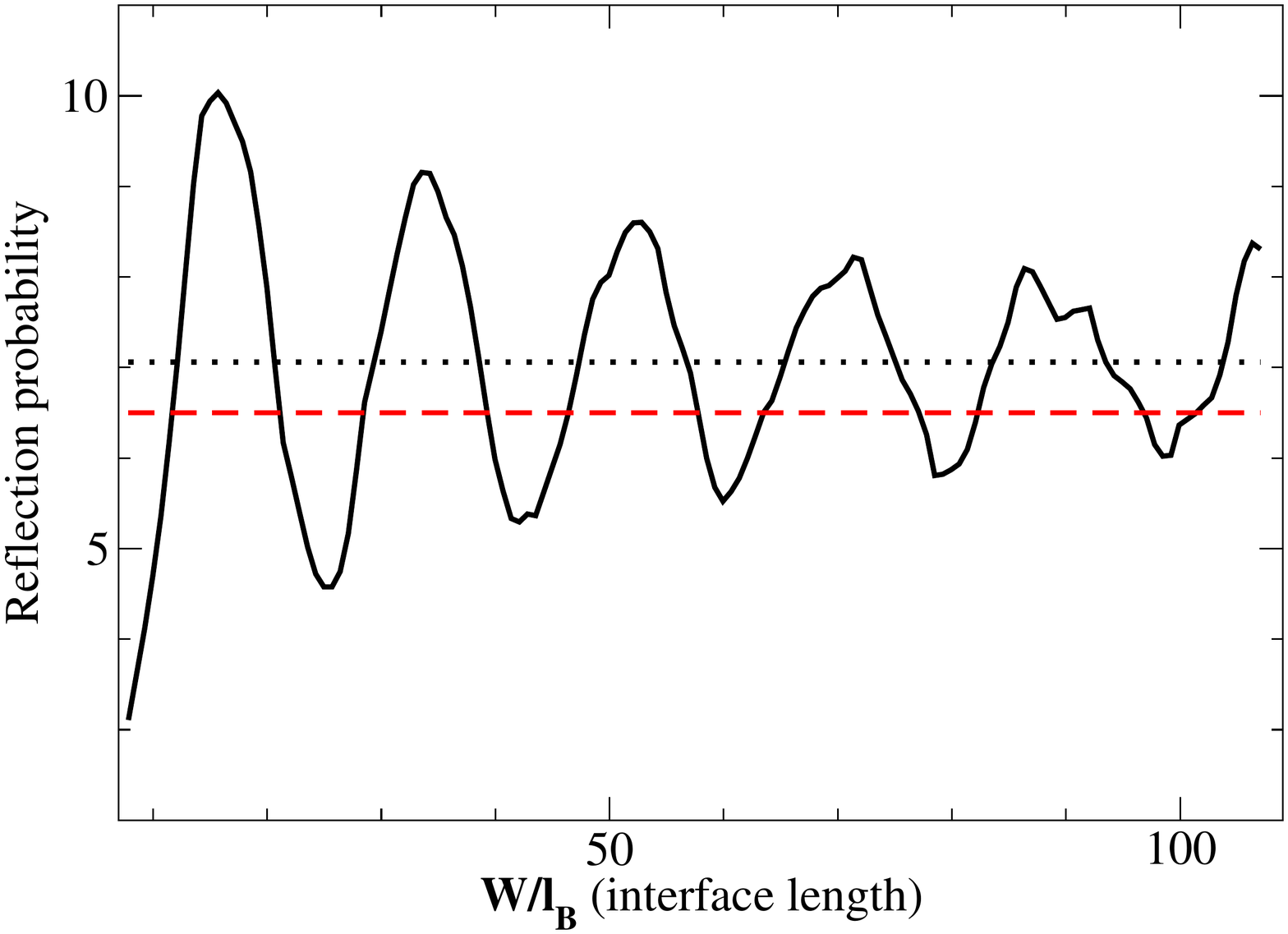}
\includegraphics[width=0.8\linewidth]{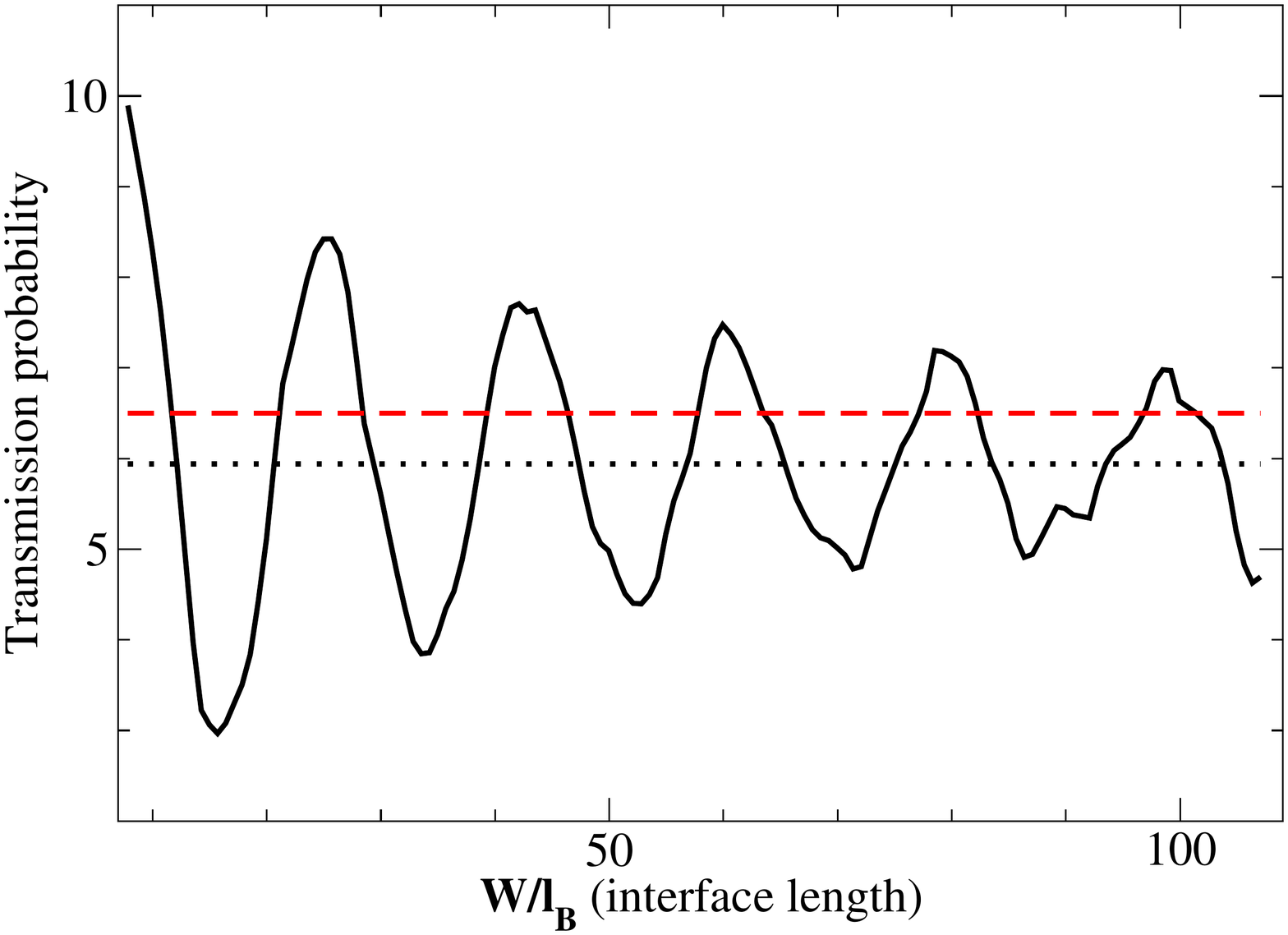}
\caption{(Color online) Reflection probability $R$ (top) and transmission 
  probability $T$ (bottom) as a function of the length of the interface $W$ 
  (in units of the magnetic length $l_B$). The electrostatic potentials are 
  $V_2 = -V_1 = 3.57$ (in units of $\epsilon_B = \hbar v_F/l_B$), which 
  corresponds to filling ratios $\nu_1 = \nu_2 = (V_2/\epsilon_B)^2/2 = 6.37$, 
  and thus 13 channels (7 states). The mean value of these functions 
  (dotted, black online) differs from the classical limit 
  (dashed, red online). The reflection and transmission probabilities of each 
  channel behave in a similar fashion, oscillating strongly as a function of 
  $W$. The variance of these oscillations is not expected to diminish 
  asymptotically as $W \to +\infty$ (see text).}
\label{FigScattering}
\end{center}
\end{figure}

\begin{figure}
\begin{center}
\includegraphics[width=0.8\linewidth]{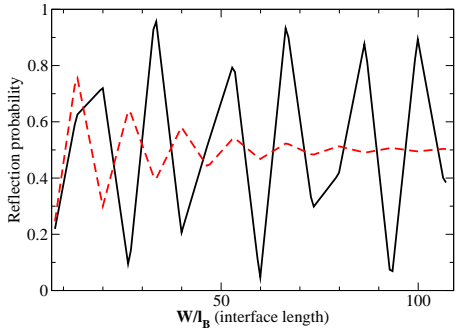}
\caption{(Color online) Semiclassical (solid, black online) and classical 
  (dashed, red online) reflection probabilities, as a function of $W$ (in 
  units of $l_B$), for the channel $(n,\alpha) = (2,+)$ (which has Klein 
  transmission probability $T(\theta_{2,+}) = 0.87$). The characteristics 
  of the ribbon are the same as in Fig.~\ref{FigScattering}. Note that the 
  semiclassical curves are locally straight lines due to the piecewise 
  linear form of the reflection and transmission probabilities in 
  Eq.~(\ref{ScattSol}).}
\label{FigCompar}
\end{center}
\end{figure}

The total reflection and transmission probabilities $R = \sum_{n,\alpha} R_{n,\alpha}$ and $T = \sum_{n,\alpha} T_{n,\alpha}$ evaluated in this way as a function of interface length are plotted on Fig.~\ref{FigScattering}.
It shows large oscillations with no sign of emergence of an asymptotic behavior (and of course no conductance plateaus). More unexpectedly, the mean value of the semiclassical curves in Fig.~\ref{FigScattering} differs from the classical limit. This is also directly visible when comparing classical and semiclassical behaviors of individual channels as in Fig.~\ref{FigCompar}. It unambiguously signals that interferences between trajectories at the potential interface dominate the physics here.

\begin{figure}
\begin{center}
\includegraphics[width=0.9\linewidth]{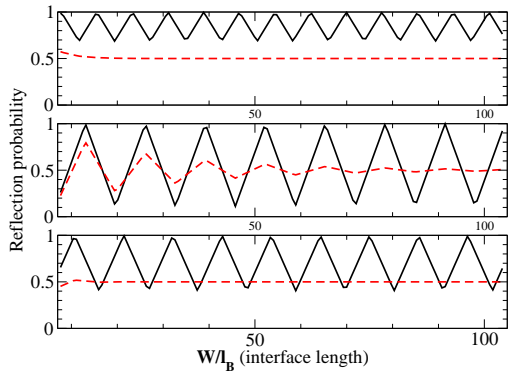}
\caption{(Color online) Semiclassical (solid, black online) and classical 
  (dashed, red online) reflection probabilities for the individual channels 
  $(n,\alpha) = (0,-),\, (2,+)$ and $(5,-)$. The ribbon considered has a 
  filling factor $\nu=6$. 
  The filling factor $\nu$ being an integer, this configuration is 
  characterized by total reflection of each channel every even bounce 
  and by mean values significantly different from the classical limit, 
  especially for small or large angles $\theta_{n,\alpha}$, as is illustrated 
  in top and bottom charts. For intermediate angles (middle chart), the 
  deviation is less pronounced but still noticeable (compare this chart with 
  Fig.~\ref{FigCompar}).}
\label{NewFig1}
\end{center}
\end{figure}

The semiclassical behavior can be understood rather straightforwardly from the scattering matrix picture. The matrix (\ref{ScattSemi}) can indeed be interpreted as that of a rotation operator on the Bloch sphere acting on vector $Z_i$, defining in this way a discrete map on the Bloch sphere. Writing this rotation operator in the form $R_{\bf n}(\omega) = \cos{(\omega/2)} 1\!\!1 - i\sin{(\omega/2)} {\bf n}\cdot\vec{\sigma}$, where ${\bf n}$ is the rotation axis on the Bloch sphere and $\omega$ the rotation angle, and identifying with the unitary matrix in Eq.~(\ref{ScattSemi}), one gets $\omega = 2\phi$ and 
\begin{equation}
\label{Axis}
{\bf n} = \frac{1}{\sin{\phi}} \left( \begin{array}{l}
    -\sin{\theta_{n,\alpha}}\sin{(\pi\nu)} \\
    -\sin{\theta_{n,\alpha}}\cos{(\pi\nu)} \\
    \cos{\theta_{n,\alpha}}\cos{(\pi\nu)} \end{array} \right) \; . 
\end{equation}
For an incoming electron (whose initial Bloch vector $Z_0$ points at the north pole), each excursion along the interface thus amounts to a rotation, on the Bloch sphere, of angle $\omega$ and around the axis given by Eq.~(\ref{Axis}). Two limiting cases furthermore provide us with a particularly simple picture. Indeed, if the filling factor is an integer, i.e.\ $\nu=n_{\rm max}$\,\footnote{Note however that in that case, and as already mentioned in section~\ref{secMag}, the quantization condition Eq.~(\ref{quantS1}) is not applicable for $n=n_{\rm max}$ and a uniform approximation should be used there.} we have
\begin{equation}
\label{eq:int_nu}
{\bf n} =  
\left( \begin{array}{l}
    0 \\ -\sin{\theta_{n,\alpha}}\\     \cos{\theta_{n,\alpha}}
            \end{array} \right) \qquad \omega = \pi 
\qquad \mbox{[integer $\nu$]} \; . 
\end{equation}
In that case the axis of rotation depends on $\theta_{n,\alpha}$, being close to $\hat {\bf z}$ for the smallest ($0,1,\cdots$) and the largest ($n_{\rm max}, n_{\rm max}\!-\!1,\cdots$) channel numbers, and near the equator for $n \simeq n_{\rm max} /2$. On the other hand, the angle of rotation is $\pi$ for everybody, implying in particular that there is total reflection for an even number of bounces. This is illustrated on Fig.~\ref{NewFig1}, where we observe that for the integer-$\nu$ case considered here, the mean value of the transmission and reflection differ significantly from the classical $1/2$ value.

\begin{figure}
\begin{center}
\includegraphics[width=0.8\linewidth]{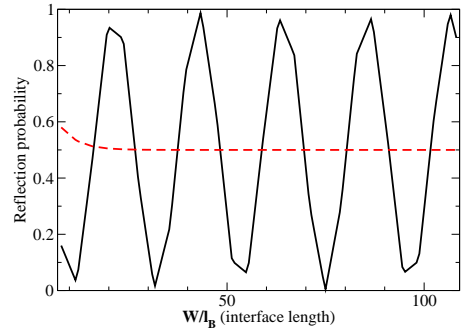}
\caption{(Color online) Semiclassical (solid, black online) and classical 
  (dashed, red online) reflection probabilities of the lowest channel for 
  a half-integer filling factor $\nu=6.5$. 
  This case is characterized by an axis of rotation $\bf n$ of 
  $R_{\bf n}(\omega)$ lying in the equatorial plane and therefore the mean 
  value of the reflection probability coincides with the classical limit 
  (to be compared with the upper chart of Fig.~\ref{NewFig1}). Also note 
  the estimated wavelength of the oscillation between reflection and 
  transmission $\lambda_\nu=\pi\sqrt{2\nu}\simeq11$ (in units of $l_B$) is 
  observable on this plot.}
\label{NewFig2}
\end{center}
\end{figure}

If, on the other hand, $\nu = n_{\rm max} + 1/2$ lies midway between two integers, we have 
\begin{equation}
\label{eq:halfint_nu}
{\bf n} =  
\left( \begin{array}{l}
    -1 \\  0 \\    0
            \end{array} \right) \qquad \omega = 2  \theta_{n,\alpha}
\qquad \mbox{[half-integer $\nu$]} \; . 
\end{equation}
The axis of rotation is then ${\bf n} = -\hat {\bf x}$, and is thus within the equator and independent on $\theta_{n,\alpha}$. As a consequence, the mean value of of the transmission and reflection coefficients will correspond in that case (and in that case only) to the classical value 1/2. The angle of rotation is now however $(\theta_{n,\alpha})$-dependent and [modulo $(2\pi)$] is close to zero for the smallest and the largest channel numbers, and close to $\pi$ for $n \simeq n_{\rm max} /2$. One should bear in mind however that $L_1=2R_1 \sin(\theta_{n,\alpha})$. The total rotation $\omega_{\rm tot} = N \omega$, where $N \simeq W/L_1$ is the number of excursions necessary to cross the junction, is thus such that
\begin{equation}
\label{eq:omega_tot}
\omega_{\rm tot} \simeq \frac{1}{\sqrt{2 \nu}} \frac{W}{l_B}
\frac{\theta_{n,\alpha}}{\sin(\theta_{n,\alpha})}  \; .
\end{equation}
For small or large channel numbers, the last factor is of order one and thus, if the width of the junction is measured in units of $l_B$, the wavelength of the oscillation between transmission and reflection as a function of $W/l_B$ is $\lambda_\nu = \pi \sqrt{2\nu}$, which is indeed what is observed on Fig.~\ref{NewFig2}.  
This wavelength is reduced by a factor $\simeq \pi/2$ for intermediate values $n \simeq \nu/2$.

\begin{figure}
\begin{center}
\includegraphics[width=0.8\linewidth]{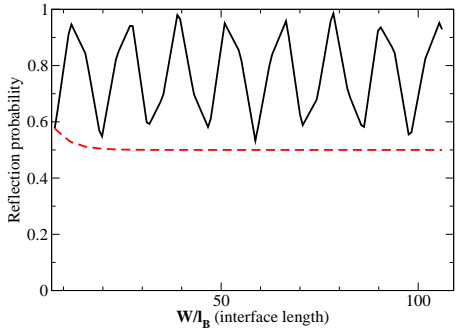}
\caption{(Color online) Semiclassical (solid, black online) and classical 
  (dashed, red online) reflection probabilities of the lowest channel for 
  a ``generic'' filling factor (i.e.\ neither integer nor half-integer) $\nu=6.25$. 
  Comparing this plot with Fig.~\ref{NewFig2} and with the upper chart of 
  Fig.~\ref{NewFig1}, we see that this case is an intermediate situation 
  in terms of deviation of the mean value from the classical limit.}
\label{NewFig3}
\end{center}
\end{figure}

Finally Fig.~\ref{NewFig3} illustrates an intermediate situation ($\nu = n_{\rm max} + 1/4$) for which both ${\bf n}$ and $\omega$ are $(\theta_{n,\alpha})$-dependent. We find a semiclassical reflection which is in this case lower than in the integer case, but still noticeably larger than the classical 1/2 value, as well as a wavelength for the oscillation between consecutive reflections which scales as $2\pi \sqrt{2\nu}$.

In summary, the classical transmission through a symmetric $n$-$p$ junction in the electric regime $\beta\ll1$ coincides with that of Eq.\ \eqref{RMTfmm}. Coherent transport through the $n$-$p$ interface gives rise to interference effects that are characterized by quantum fluctuations which depend on $W/l_B$ and $\nu$. Those fluctuations are different than the UCF predicted by RMT.

The symmetric case corresponds to the special situation where the doping is the same at both $n$ and $p$ regions. There is no {\sl a priori} reason to expect the transmission to be the same for both $V_2=-V_1$ and $V_2 \neq -V_1$. Hence, to make contact with experiments, we need a theory for the general case where the gate voltages are not symmetric. This is what we do next.

\section{Electric regime: general case}
\label{secElecGen} 

This section is devoted to the calculation of the transmission through a $n$-$p$ junction for the general case, where $V_1\neq -V_2$, in the electric regime. For that purpose, it is no longer possible to use the intuitive scattering matrix approach discussed in the previous section, which is valid for the symmetric case only. In what follows we develop a semiclassical theory that allows the calculation of the Landauer transmission for the general case. We start discussing how to apply the Fisher-Lee formula \cite{Fisher81} for a graphene $n$-$p$ junction. Then we address the dynamics at the interface, which provides the elements required by the semiclassical calculation. We proceed presenting the main technical details of the derivation, complemented by some additional material presented in the appendices. We conclude this section discussing the classical limit and summarizing the main features of the transmission in the general case.

\subsection{Fisher-Lee/Baranger-Stone formalism}

We now consider the electric regime of a step-like junction for arbitrary values of $V_1$ and $V_2$ ($V_2V_1 < 0$), for which we obtain a semiclassical evaluation of the Landauer conductance, Eq.\ (\ref{LanButt}). This is achieved with the help of a formalism which was first introduced by Fisher and Lee \cite{Fisher81}, and later generalized to account for a magnetic field by Baranger and Stone \cite{Baranger89}. This formalism is based on the use of Green's functions for which we derived in a previous work \cite{Carmier08} a semiclassical approximation in graphene. As was discussed in that paper, the distinguishing feature of this Green's function as compared with the standard 2DEG Schr\"odinger expression is the appearance of a semiclassical phase which can be understood as the topological part of the usual Berry phase occurring in the context of systems depending adiabatically on an external parameter. However in the absence of a mass term in the graphene Hamiltonian (as will be the case in this work), both phases are equal and can be expressed as $\xi_{sc} = -(\theta'' - \theta')/2$, with $\theta'$ and $\theta''$ the angle of the initial and final pseudo-momentum $\Pi$ of the corresponding trajectory. We shall thus refer to it in the following as the Berry phase.



Turning back to the Fisher-Lee/Baranger-Stone formalism, the formulae obtained in [\onlinecite{Fisher81,Baranger89}] were derived for Schr\"odinger (scalar) electrons and should be somewhat modified to describe charge carriers in graphene. Special attention must be paid to the pseudo-spin degree of freedom which shows up in the spinor structure of the modes $\chi_{n,\alpha}$ and the matrix structure of the Green's function and which generates non-commutative operations. With this in mind, calculations are rather straightforward and the following expressions can be obtained for a general mesoscopic graphene sample with an arbitrary number of leads ($\mu$,$\nu$,$\cdots$): the conductance from lead $\nu$ to lead $\mu$ (with $\mu \neq \nu$) is
\begin{equation}
\label{gmunu}
\begin{split}
g_{\mu\nu} = - \frac{e^2 \hbar}{2\pi} \int_{{\cal C}_{\mu}} d y_{\mu}
\int_{{\cal C}_{\nu}} d y'_{\nu} Tr \left[ \left(\nabla_{\hat{\bf p}}
    {\cal H} \cdot {\bf e}_{\mu}\right) \right.
\\
\times \left. {\cal G}({\bf r}_{\mu},{\bf
    r'}_{\nu}) \left(\nabla_{\hat{\bf p}} {\cal H} \cdot {\bf
      e}_{\nu}\right) {\cal G}^{\dagger}({\bf r}_{\mu},{\bf r'}_{\nu})
\right] 
\end{split}
\end{equation}
and the transmission probability amplitude of going from channel $n$ in lead $\nu$ to channel $m$ in lead $\mu$ is 
\begin{equation}
\label{tmunumn}
\begin{split}
t_{\mu\nu,mn} = -i\hbar \int_{{\cal C}_{\mu}} d y_{\mu} \int_{{\cal
    C}_{\nu}} d y'_{\nu} \chi^{+\dagger}_m ({\bf r}_{\mu}) \left(
  \nabla_{\hat{\bf p}} {\cal H} \cdot {\bf e}_{\mu} \right)
\\
\times {\cal
  G}({\bf r}_{\mu},{\bf r'}_{\nu}) \left( \nabla_{\hat{\bf p}} {\cal
    H} \cdot {\bf e}_{\nu} \right) \chi^{-}_n ({\bf r'}_{\nu}) \; . 
\end{split}
\end{equation}
With these notations, the Landauer-B\"uttiker formula reads $g_{\mu\nu} = G_0\sum_n\sum_m |t_{\mu\nu,mn}|^2$. ${\cal C}_\mu$ and ${\cal C}_\nu$ are transverse sections of the leads, while ${\bf e}_\mu$ and ${\bf e}_\nu$ stand for the unit normal (outward pointing) vectors to the corresponding  sections. ${\cal H}$ is the graphene Hamiltonian, ${\cal G}$ the (retarded) Green's function and $\chi_n^\pm$ the quantized mode (with $\pm$ labelling its direction of propagation with respect to the central region separating the leads). Dependence on the valley index $\alpha$ in the modes has been temporarily dropped for convenience.

The expressions (\ref{gmunu}) and (\ref{tmunumn}) can be slightly lightened when the two valleys $K$ and $K'$ are uncoupled and can be treated independently. Starting from the valley isotropic representation ${\bf \Psi} = (\psi_A,\psi_B,-\psi'_B,\psi'_A)^T$ introduced in section \ref{secMag}, the effective Hamiltonian within the valley $\alpha$ can be written
\begin{equation}
{\cal H} = v_F \hat{{\bf \Pi}}\cdot\vec{\sigma} + U({\bf r}) 1\!\!1 +
\alpha m({\bf r}) v_F^2 \sigma_z \; ,
\end{equation}
with the convention that
\begin{equation}
\label{PsiIso}
\Psi = \left( \begin{array}{l} \psi_1 \\ \psi_2 \end{array} \right) =
\left| \begin{array}{l} (\psi_A,\psi_B)^T  \;\;\; \text{if} \;\;
    \alpha = 1 \vspace*{0.2cm} \\ (-\psi'_B, \psi'_A)^T \;\;\;
    \text{if} \;\; \alpha = -1 \end{array} \right. \; . 
\end{equation}
With this choice of representation, expressions (\ref{gmunu}) and (\ref{tmunumn}) read 
\begin{equation}
\begin{split}
g_{\mu\nu}^{\mu \neq \nu} = - G_0\frac{(\hbar v_F)^2}{2}
\int_{{\cal C}_{\mu}} d y_{\mu} \int_{{\cal C}_{\nu}} d y'_{\nu} Tr
\left[ \left(\vec{\sigma} \cdot {\bf e}_{\mu}\right) \right.
\\
\times \left. {\cal G}({\bf
    r}_{\mu},{\bf r'}_{\nu}) \left(\vec{\sigma} \cdot {\bf
      e}_{\nu}\right) {\cal G}^{\dagger}({\bf r}_{\mu},{\bf r'}_{\nu})
\right] \; , 
\end{split}
\end{equation}
\begin{equation}
\begin{split}
t_{\mu\nu,mn} = -i\hbar v_F^2 \int_{{\cal C}_{\mu}} d y_{\mu}
\int_{{\cal C}_{\nu}} d y'_{\nu} \chi^{+\dagger}_m ({\bf r}_{\mu}) \left( \vec{\sigma} \cdot {\bf e}_{\mu} \right)
\\
\times
{\cal G}({\bf r}_{\mu},{\bf r'}_{\nu}) \left( \vec{\sigma} \cdot {\bf e}_{\nu}
\right) \chi^{-}_n ({\bf r'}_{\nu}) \; . 
\end{split}
\end{equation}

Focusing now on the specific geometry under consideration (cf. Fig.~\ref{FigBetPoin}), we can drop the lead indices, assume the sections from which the conductance is computed to be located at the extremities of the junction (at abscissa $x'$ on the incoming side and $x''=x'+W$ on the outgoing one) and use the coordinate $y$ inside the section. The transmission coefficients can then be written as
\begin{equation}
\label{tmnribbon}
t_{mn} = i\hbar v_F^2 \int dy'' \int dy' \chi_m^{+\dagger}({\bf r''})
\sigma_x {\cal G}({\bf r''}, {\bf r'}) \sigma_x \chi_n^{-}({\bf r'})
\; . 
\end{equation}
Note that since we are working in the quantum Hall regime $W \gg l_B$, integrals in Eq.~(\ref{tmnribbon}) are effectively restricted to one of the edges. 

As we are interested in the total conductance rather than the individual transmission coefficients, we do not need to compute all the $t_{mn}$ but only the sum $\sum_m |t_{mn}|^2$. Using that 
\begin{equation} \label{eq:normalization}
\int dy \chi^{\pm\dagger}_m ({\bf r}) v_F \sigma_x \chi^{\pm}_n
({\bf r}) = \delta_{mn}
\end{equation}
as is proven in \cite{Baranger89}, one can easily show that
\begin{equation}
\label{Tnalpha}
T_{n,\alpha} \stackrel{def}{=} \sum_m |t_{mn}|^2 = \int_{y''>0} dy'' {\cal
  T}_{n,\alpha}^{\dagger}({\bf r''}) v_F \sigma_x {\cal
  T}_{n,\alpha}({\bf r''}) 
\end{equation}
with
\begin{equation}
\label{calTnalpha}
{\cal T}_{n,\alpha}({\bf r''}) = i\hbar\int dy' {\cal G}({\bf r''},
{\bf r'}) v_F \sigma_x \chi_{n,\alpha}^{-}({\bf r'}) \; .
\end{equation}
The same expressions apply for $R_{n,\alpha} = 1 - T_{n,\alpha}$, except that in Eq.~(\ref{Tnalpha}) the integral should be taken in the electron side of the junction, i.e. on $y''<0$. The prescription of directly computing $T_{n,\alpha}$ instead of the individual $t_{mn}$ additionally bypasses the need to project the incoming modes $\chi_n^-$ propagated along the interface on the outgoing ones $\chi_m^+$ in Eq.~(\ref{tmnribbon}). This is particularly useful on the $p$ side of the junction where angle $\theta_2$ has no reason to coincide with a quantized value and where transmitted charge carriers are therefore no longer in a properly quantized state but in a superposition of outgoing modes.

Our main task is now to evaluate semiclassically Eq.~(\ref{calTnalpha}). This requires obtaining semiclassical approximations of the incoming mode $\chi_{n,\alpha}^-({\bf r'})$ and of the Green's function ${\cal G}({\bf r''},{\bf r'})$. The mode $\chi_{n,\alpha}^-({\bf r'})$ is built semiclassically on the manifold obtained from the one-parameter family of trajectories bouncing with an angle $\theta_{n,\alpha}$ on the edge of the lead. Within the representation (\ref{PsiIso}) and sticking with an infinite mass edge confinement, one gets
\begin{equation}
\label{modesemi}
\begin{split}
\chi_{n,\alpha}^-({\bf r}) = \frac{C_{n,\alpha} e^{i k_x^{n,\alpha}
    x}}{\sqrt{|\sin{\theta_{n,\alpha}(y)}|}} \sum_{\nu = \pm 1}
e^{\frac{i}{\hbar}\nu S_{n,\alpha}(y) + i\frac{\pi}{2}\mu(\nu)}
\\
\times
\left( \begin{array}{l} e^{-i\frac{\nu}{2}\theta_{n,\alpha}(y)} \\
    e^{i\frac{\nu}{2}\theta_{n,\alpha}(y)} \end{array} \right) +
O(\hbar) \; ,
\end{split}
\end{equation}
with $\nu=\pm$ an index (not to be confused with the filling factors $\nu_i$) labeling the sheets of the manifold on which the mode is constructed ( $p_y > 0$ for $\nu = +1$, $p_y < 0$ for $\nu = -1$). The caustic in phase space at the junction of the two sheets is taken into account by the phase jump $\mu(\nu) = \Theta(-\nu)$ (with $\Theta$ the Heaviside step function). In Eq.~(\ref{modesemi}) $\theta_{n,\alpha}(y) = \cos^{-1}{(\cos{\theta_{n,\alpha}} - y/R_1)}$ is the angle of the tangent to the trajectory when at a distance $y$ from the edge, $k_x^{n,\alpha} = k_F \cos{\theta_{n,\alpha}}$ is the constant of motion associated with the mode, $S_{n,\alpha}(y) = \hbar\nu_1f(\theta_{n,\alpha}(y))$ the action, and $C_{n,\alpha} = (4v_FR_1\sin{\theta_{n,\alpha}})^{-1/2}$ is a normalization constant which is determined from Eq.~(\ref{eq:normalization}).

Turning now to the Green's function, a semiclassical approximation valid in either the electron or hole region was derived in \cite{Carmier08}. Including Klein tunneling, i.e.\ the transitions from electron to hole regions to this formalism, is a priori a non-trivial (although feasible) task in a completely general setup. The limit $\beta \ll 1$ that we consider here, and the fact that we assume a perfectly straight potential step, simplify however considerably the problem. Indeed one can in this case, for the Klein tunneling, treat the semiclassical wavefunctions as plane waves, and therefore use the transmission and reflection coefficients Eqs.~(\ref{eq:r1})-(\ref{eq:t1}) as in \cite{Couchman92}. This leads to the straightforward generalization of the semiclassical Green's function expression 
\begin{equation}  \label{eq:Gsc}
\begin{split}
{\cal G}_{sc}&({\bf r''},{\bf r'};\epsilon_F) = \sum_j \left(
  \prod_{i=1}^2 \prod_{\gamma_i,\eta_i} r_i^{(\gamma_i)}
  t_i^{(\eta_i)} \right) 
\\
& \times \frac{e^{\frac{i}{\hbar}S_j({\bf r''},{\bf
      r'}) - i\frac{\pi}{2}\mu_j + i\xi_j}}{i\hbar\sqrt{2\pi
    i\hbar|J_j({\bf r''},{\bf r'})|}} V_j^{\epsilon({\bf r''})}({\bf
  r''}) V_j^{\epsilon({\bf r'})\dagger}({\bf r'}) \; , 
\end{split}
\end{equation}
with $\gamma_i,\eta_i$ labelling scattering events on the $n$-$p$ junction, respectively associated to reflections and transmissions in region $i$. As usual, the sum is over all classical trajectories joining points ${\bf r'}$ and ${\bf r''}$ at the Fermi energy $\epsilon_F$. The phases accumulated along the way include the action $S_j$, the Maslov index $\mu_j$ counting the number of caustics traversed, and the Berry phase $\xi_j = -(\theta''_j - \theta'_j)/2$, with $\theta'_j$ and $\theta''_j$ the direction of the initial and final pseudo-momentum $\bf \Pi$ of the trajectory $j$. Note that as the total Berry phase is accounted for in $\xi_j$, the reduced reflection and transmission coefficients (without the Berry phases) $r_i$ and $t_i$ given by (\ref{eq:r1})-(\ref{eq:t1}) should be used in Eq.~(\ref{eq:Gsc}). The determinant
\begin{equation} \label{eq:J}
J_j({\bf r''},{\bf r'}) = -\dot{x}_j''\dot{x}_j'
\left( \frac{\partial^2 S_j}{\partial y'' \partial   y'} \right)^{-1}
\end{equation}
implements the conservation of classical probability. Finally, $V_j^{\epsilon({\bf r})}({\bf r})$ is the eigenstate along the $j^{th}$ trajectory of the classical Hamiltonian $H^{\epsilon({\bf r})} = V(x) + \epsilon({\bf r})v_F|{\bf \Pi}|$, with $\epsilon({\bf r}) = \pm 1$ ($+1 $ on the electron side and $-1$ on the hole side). In the absence of any mass term in the bulk of the sample and choosing the representation (\ref{PsiIso}), these eigenstates depend solely on the angle of the pseudo-momentum: $V_j^+({\bf r}) = (1/\sqrt{2}) (1,e^{i\theta_j})^T$ and $V_j^-({\bf r}) = (1/\sqrt{2}) (e^{-i\theta_j},-1)^T$.

Determining the specific Green's function for the problem under consideration can essentially be reduced to the task of making a complete list of the trajectories connecting the Poincar\'e sections on both sides of the interface and computing the probability amplitudes associated to each one of them. This issue will now be addressed.

\subsection{Dynamics at the interface}

Depending on the relative size of the potentials $|V_1|$ and $V_2$, Eq.~(\ref{conserv}) defines a critical angle for either $\theta_1$ or $\theta_2$ above which reflection on the $n$-$p$ junction is total. Without loss of generality, we will take $|V_1| < V_2$ which constrains incident holes on the interface to the angular domain [$\theta_{crit}$,$\pi - \theta_{crit}$], with $\theta_{crit} = \cos^{-1}{(|V_1|/V_2)}$. This allows not to worry about possible total reflection on the electron side, and additionally sets the excursion length scales $L_2 > L_1$.

\begin{figure}
\begin{center}
\includegraphics[width=0.99\linewidth]{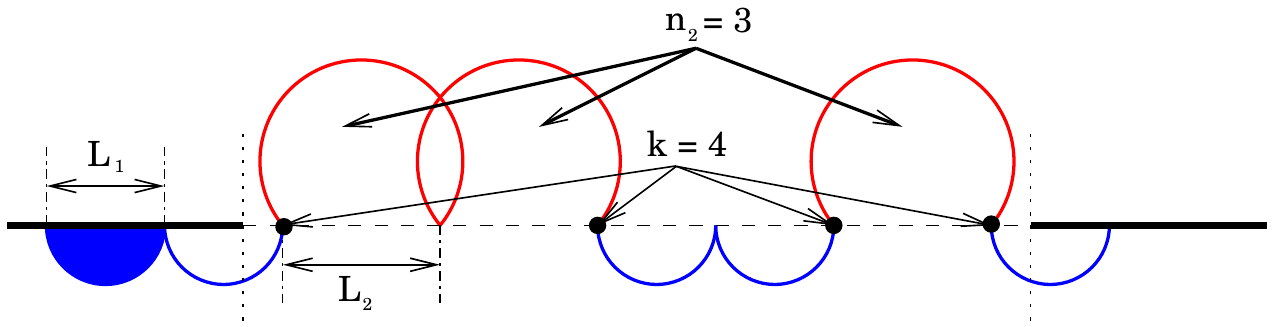}
\caption{(Color online) Trajectory characterized by $n_2 = 3$ excursions 
  on the hole side and $k = 4$ traversals of the interface} 
\label{FigGenSout}
\end{center}
\end{figure}

Consider a typical trajectory going from ${\bf r'} = (x',y')$ in the initial $x=x'$ Poincar\'e section to ${\bf r''} = (x'',y'')$ in the final $x = x''=x'+W$ Poincar\'e section. The trajectory can be labeled by an index $j$ specifying whether the trajectory is transmitted or reflected for each of its successive encounters with the junction. For a given $j$, the initial and final angles $\theta'$ and $\theta''$ are fixed once the coordinates $y'$ and $y''$ are. The trajectory $j$, that is the successive list of transmissions and reflections, can be characterized by two integers: the number of excursions $n_2$ in region 2 (which then fixes the remaining number $n_1(n_2)$ of excursions in region 1) and the number of traversals $k$ of the interface. A typical example is shown in Fig.~\ref{FigGenSout}. 

These two integers do not specify uniquely the trajectory $j$ since it is possible to permute the order of the excursions in the hole and electron regions while preserving the couple ($n_2$,$k$). They however define a class of trajectories which, as we will see, give the same contribution to the semiclassical Green's function Eq.~(\ref{eq:Gsc}). Indeed, the mapping $(y',\theta') \mapsto (y'',\theta'')$ depends only on $n_2$, which fixes the number of excursions in regions 1 and 2, and on the parity of $k$, which determines whether the trajectory $j$ exits the junction in region 1 ($k$ even, reflection) or 2 ($k$ odd, transmission). This mapping remains however unchanged if the order of the excursions in regions 1 and 2 is modified. As a consequence the determinant Eq.~(\ref{eq:J}), which can be expressed in terms of this mapping, or the Berry phase $\xi_j = -(\theta'' - \theta')/2$ which is only a function of the initial and final angles of the trajectory, are also independent of the ordering of the excursions.  

In the same way, the action $S_j$ and the Maslov index $\mu_j$ in the phase of the semiclassical Green's function Eq.~(\ref{eq:Gsc}) are functions of $n_2$ and the parity of $k$ only. The action, for instance, can be expressed as $S_j(n_2)= \hbar k_x W + n_1(n_2)S_1 + n_2S_2 + \Delta S$, with $S_1 =2\hbar\nu_1f(\theta_1)$ and $S_2 = -2\hbar\nu_2f(\pi - \theta_2)$) the actions accumulated along an excursion in regions 1 and 2 respectively (note that $S_2 < 0$ since, in region 2, ${\bf p}\cdot\dot{\bf r} < 0$), and
\begin{equation}
\begin{split}
 \Delta S & = \hbar\nu_1 \left( f(\theta_1) - f(\theta') \right)
\\
  &  + \left| \begin{array}{l} \hbar\nu_1 \left( f(\theta'') -
        f(-\theta_1) \right) \; \;   \; \; (\mbox{$k$ even, reflection})
      \vspace*{0.2cm} \\ 
      \hbar\nu_2 \left( f(\theta'') - f(\pi - \theta_2) \right)
      \; \; (\mbox{$k$ odd, transmission}) \end{array} \right. 
\end{split}
\end{equation}
($f : \theta \to \theta - (\sin{2\theta})/2$ is the same function as in section~\ref{secMag}). The angles $\theta_1$ and $\theta_2$ are the ones introduced in Fig.~\ref{FigBetPoin} and at this point should be understood as being functions of $\theta'$ and $\theta''$. Explicit computation of the Maslov index $\mu_j(n_2)$ (see appendix~\ref{appMaslov}) shows also that, quite naturally, it does not depend either on the ordering of the excursions.

We now turn our attention to the factors associated with scattering at the interface. Let us first consider the case where the trajectory exits the junction in region 1, i.e. of an even number of traversals $k = 2k'$. In that case, the number of traversals from 1 to 2 as well as from 2 to 1 are equal to $k'$, and there are $n_2 - k'$ reflections on side 2 and $n_1(n_2) + 1 - k'$ reflections on side 1 (the additional term $1$ coming from the fact that the trajectory initially leaves the Poincar\'e section in that region). The probability amplitude associated with the reflections and transmissions at the junction interface for the class of trajectories ($n_2$,$k'$) is thus given by
\begin{equation}
\label{ARprob}
A_R(n_2,k') = r_1^{n_1(n_2) + 1}r_2^{n_2} \left( \frac{t_1t_2}{r_1r_2}
\right)^{k'} \; . 
\end{equation}
The case of an odd number of traversals $k = 2k'' + 1$, i.e. when the trajectory $j$ is eventually transmitted in region 2, is completely equivalent. There are then $k'' + 1$ transmissions from region 1 to region 2, $k''$ transmissions from region 2 to region 1, $n_2 - k''$ reflections in region 2 and $n_1(n_2) - k''$ in region 1. The probability amplitude for the class ($n_2$,$k''$) thus reads 
\begin{equation}
A_T(n_2,k'') = t_1r_1^{n_1(n_2)}r_2^{n_2} \left( \frac{t_1t_2}{r_1r_2}
\right)^{k''} \; . 
\end{equation}

\begin{figure}
\begin{center}
\includegraphics[width=0.7\linewidth]{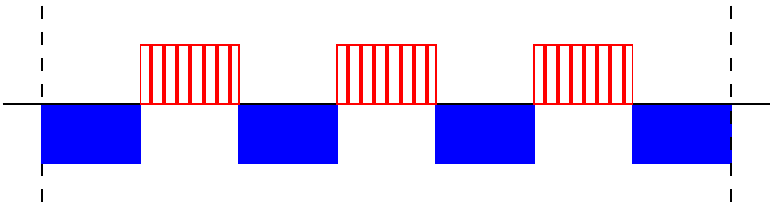}
\caption{(Color online) Class of trajectories characterized by $k' = 3$ 
  (and an arbitrary $n_2$). Filled (blue online) rectangles symbolize a 
  succession of excursions in region 1 and hatched (red online) rectangles 
  a succession of excursions in region 2. Although these rectangles are 
  represented as if they had the same size, each one of them may contain a 
  different number of excursions.}
\label{FigOmegaR}
\end{center}
\end{figure}

All permutations in the order of the excursions along the interface that preserve numbers $n_2$ and $k$ correspond to the same probability amplitude. We must therefore determine the degeneracy factor $\Omega$ that gives the number of distinct trajectories belonging to the class characterized by integers ($n_2$,$k$). Starting again with $k=2k'$ even (reflection), let us materialize each uninterrupted succession of excursions in region 1 or 2 by a rectangle (see Fig.~\ref{FigOmegaR}). 
If we include excursion portions respectively leaving from the initial Poincar\'e section and arriving at the final Poincar\'e section, there are $k'$ rectangles in region 2 and $k' + 1$ rectangles in region 1, when the total number of excursions in region 2 is $n_2$, and that in region 1 is $n_1(n_2) + 2$. Then, using the combinatorial result that there are $\dbinom{n - 1}{k - 1}$ ways of writing an integer $n$ as a sum of $k$ non-zero integers (or equivalently of distributing $n$ excursions into $k$ non-empty intervals), the degeneracy factor is straightforwardly given by
\begin{equation}
\Omega_R(n_2,k') = \dbinom{n_2 - 1}{k' - 1} \dbinom{n_1(n_2) + 1}{k'}
\end{equation}
with $\dbinom{n}{k} = \frac{n!}{k!(n - k)!}$ the binomial coefficient.

For an odd number $k=2k''+1$ of traversals (transmission), there should now be an equal number $k'' + 1$ of rectangles in regions 1 and 2, while the total number of excursions are respectively $n_1(n_2) + 1$ in region 1 and $n_2 + 1$ in region 2, once more including here ``partial'' excursions leaving from the initial Poincar\'e section and arriving at the final Poincar\'e section. Using the same combinatorial result as before, this yields for the degeneracy factor
\begin{equation}
\Omega_T(n_2,k'') = \dbinom{n_2}{k''} \dbinom{n_1(n_2)}{k''} \; .
\end{equation}

Combining all of the results obtained in this subsection, the semiclassical Green's function for our problem can be expressed as 
\begin{widetext}
\begin{equation}
\label{semiG}
{\cal G}_{sc}({\bf r''},{\bf r'}) = 
 \sum_{n_2 = 0}^{N_2} \frac{ V_{n_2}^{\epsilon({\bf r''})}({\bf r''})
   V_{n_2}^{+\dagger}({\bf    r'}) }{i\hbar\sqrt{2\pi i\hbar|J_{n_2}({\bf r''},{\bf r'})|}}
\Xi(n_2) e^{\frac{i}{\hbar}S(n_2) -  i\frac{\pi}{2}\mu(n_2) + i\xi_{n_2}}
\end{equation}
with
\begin{equation}
\Xi(n_2) = 
\left\{ \begin{array}{ll} \sum_{k'}  A_R(n_2,k') \Omega_R(n_2,k') \qquad \quad &
    \mbox{if $y''<0$ (reflection})
    \vspace*{0.2cm} \\  \sum_{k''}  A_T(n_2,k'') \Omega_T(n_2,k'') &
    \mbox{if $y''>0$ (transmission)} \end{array} \right. 
\end{equation}
\end{widetext}
and $N_2$ the upper bound of the number of excursions in region $2$ (bounds for the number of traversals are given by $\text{min}(1,n_2) \leq k'  \leq \text{min}(n_2, n_1(n_2) + 1)$ and $0 \leq k'' \leq \text{min}(n_2, n_1(n_2))$.

\subsection{Semiclassical expression for the conductance}

Let us first discuss the case of reflected trajectories ($y''<0, \epsilon({\bf r''}) > 0$). Using the semiclassical expressions (\ref{modesemi}) and (\ref{semiG}), the matrix structure in integral (\ref{calTnalpha}) reads
\begin{equation} \label{eq:matrix_struct}
\begin{split}
V_j^{+}({\bf r''}) V_j^{+\dagger}({\bf r'}) \sigma_x
\left( \begin{array}{c} e^{-i\frac{\nu'}{2}\theta_{n,\alpha}(y')} \\
    e^{i\frac{\nu'}{2}\theta_{n,\alpha}(y')} \end{array} \right) =
e^{-i\theta_j'/2} 
\\
\times \cos{\left(\frac{\nu'\theta_{n,\alpha}(y') +
    \theta_j'}{2}\right)} \left( \begin{array}{c} 1 \\
    e^{i\theta_j''} \end{array} \right) \; . 
\end{split}
\end{equation}
Computing the integral (\ref{calTnalpha}) in the semiclassical limit $\hbar \to 0$ can be done using the stationary phase approximation $\int A(x) e^{\frac{i}{\hbar}S(x)} d x \simeq \sum_{x_s} A(x_s)e^{\frac{i}{\hbar}S(x_s)} (2\pi i\hbar/|{\partial^2_x S}(x_s)|)^{1/2} e^{-i\frac{\pi}{2}\mu_S}$, with $x_s$ the points where $S(x)$ is stationary and $\mu_S = \Theta [ -\partial^2 S/\partial x^2 (x_s)]$. The stationary phase condition
\begin{equation}
\label{statphas}
\left. \frac{\partial (\nu' S_{n,\alpha} + S_j)}{\partial y'}
\right|_{y'',n_2} = 0 \;\; \Rightarrow \; \nu'\theta_{n,\alpha}(y'_s)
= \theta'_j(y'_s) 
\end{equation}
expresses that for a given final position $y''$ (and a given number of excursions in region $2$), the stationary phase point $y'_s$ is the one where the initial angle matches with that of the quantized mode. This implies the additional identifications $\theta_1 = \theta_{n,\alpha}$, $k_x^j = k_x^{n,\alpha}$ and $\theta''_j = \nu'' \theta_{n,\alpha}(y'')$ with $\nu''$ the sheet index in the final Poincar\'e section. The action of the Green's function thus becomes $S_j(y'_s) = \hbar k_x^{n,\alpha}(x'' - x') + (n_1 + 1)S_1^{n,\alpha} + n_2S_2 + \nu'' S_{n,\alpha}(y'') - \nu' S_{n,\alpha}(y'_s)$, with $S_1^{n,\alpha} = 2\hbar\nu_1 f(\theta_{n,\alpha})$ quantized as in Eq.~(\ref{quantS1}). Inserting these results in the integral (\ref{calTnalpha}), and bearing in mind that the stationary phase point $y'_s$ depends on the integer $n_2$ we obtain
\begin{widetext}
\begin{equation}
\label{interR}
\begin{split}
{\cal T}_{n,\alpha}({\bf r''}) = v_F C_{n,\alpha}e^{i k_x^{n,\alpha}
  x''} \sum_{\nu'' = \pm 1} \sum_{n_2=0}^{N_2}
\frac{e^{\frac{i}{\hbar}\nu'' S_{n,\alpha}(y'')}
  e^{i\frac{\pi}{2}(\mu(\nu') - \mu_{J,S})}
  \cos{\theta_{n,\alpha}(y'_s)}}{\sqrt{\left| \left. {\partial^2_{y'}
          (S_j + S_{n,\alpha})} \right|_{y''} (y'_s) J_j(y'_s)
      \sin{\theta_{n,\alpha}(y'_s)} \right|}} 
\\
\times \left( \begin{array}{c}
    e^{-i\frac{\nu''}{2}\theta_{n,\alpha}(y'')} \\
    e^{i\frac{\nu''}{2}\theta_{n,\alpha}(y'')} \end{array} \right)
e^{\frac{i}{\hbar}(n_1 + 1)S_1^{n,\alpha}} e^{\frac{i}{\hbar}n_2S_2}
\sum_{k'} A_R(n_2,k') \Omega_R(n_2,k') \; . 
\end{split}
\end{equation}
\end{widetext}
In Eq.~(\ref{interR}), $\mu_{J,S}$ is the sum of the Maslov index $\mu_J$ in the Green's function and of the index $\mu_S$ coming from the stationary phase integral. The latter is zero if ${\partial^2_{y'} (S_j + S_{n,\alpha})} > 0$ and one if ${\partial^2_{y'} (S_j + S_{n,\alpha})} < 0$, while the former requires some care to be computed precisely. The technical calculation of $\mu_J$ is detailed in appendix \ref{appMaslov}. For our current purposes, one can
actually show that $\mu_{J,S} = \mu(\nu') - \mu(\nu'') + n_1 + 1 - n_2$ with $\mu(\nu'') = \Theta(-\nu'')$ the phase jump at the caustic between the sheets in the final Poincar\'e section. 

\begin{figure}
\begin{center}
\includegraphics[width=0.7\linewidth]{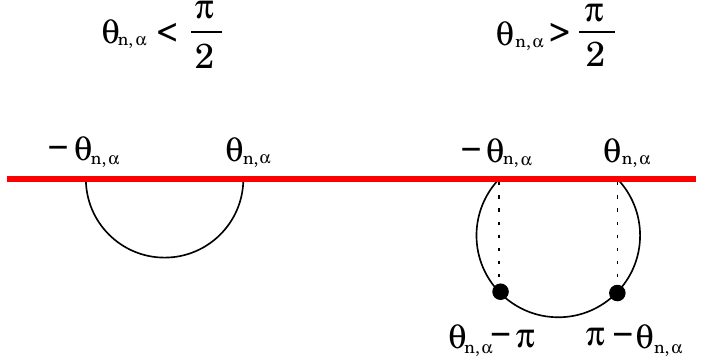}
\caption{(Color online) Portion of cyclotronic orbit illustrating that, 
  when $\theta_{n,\alpha} > \frac{\pi}{2}$, accessible values for 
  $\theta''$ do not any more correspond to the range 
  $[-\theta_{n,\alpha},\theta_{n,\alpha}]$.}
\label{FigIntBounds}
\end{center}
\end{figure}

The final step of this calculation involves computing the prefactor $(\left. {\partial^2_{y'} (S_j + S_{n,\alpha})} \right|_{y''} J_j)^{-1/2}$, which we do in appendix \ref{appPref}. Inserting the result in Eq.~(\ref{interR}), one finds that all trace of the stationary phase point $y'_s$ has vanished and that expression (\ref{interR}) can be simply written as
\begin{equation}
\label{propagTnalpha}
\begin{split}
{\cal T}_{n,\alpha}({\bf r''}) = \chi_{n,\alpha}^{+}({\bf r''})
\sum_{n_2=0}^{N_2} (-i e^{\frac{i}{\hbar}S_1^{n,\alpha}})^{n_1 + 1} (i
e^{\frac{i}{\hbar}S_2})^{n_2}
\\
\times \sum_{k'} A_R(n_2,k') \Omega_R(n_2,k') \; .
\end{split} 
\end{equation}
Comparing this with the original integral (\ref{calTnalpha}) makes it possible to give a rather transparent interpretation for the role played by the Green's function. It basically amounts to propagating the original mode from ${\bf r'}$ to ${\bf r''}$ with a certain probability weight corresponding to the various trajectories fulfilling the stationary phase condition (\ref{statphas}) and connecting these points. The reflection probability for channel $n$
polarized in valley $\alpha$ is obtained by inserting Eq.~(\ref{propagTnalpha}) in Eq.~(\ref{Tnalpha}), which immediately gives 
\begin{equation}
\begin{split}
& R_{n,\alpha} =  v_F|C_{n,\alpha}|^2\int d y''
\frac{2\cos{\theta''}}{\sin{\theta''}} 
\\
& \times \left| \sum_{n_2=0}^{N_2} (-i
  e^{\frac{i}{\hbar}S_1^{n,\alpha}})^{n_1 + 1}
 (i 
e^{\frac{i}{\hbar}S_2})^{n_2} \sum_{k'} A_R(n_2,k') \Omega_R(n_2,k')
\right|^2
\end{split} 
\end{equation}
with the shorthand $\theta'' = \theta_{n,\alpha}(y'')$. A change of variables $d y'' = R_1 \sin{\theta''} d \theta''$ leads to the final result
\begin{widetext}
\begin{equation}
\label{Rnalphafinal}
R_{n,\alpha} = \int_{-\theta_R}^{\theta_R} d \theta''
\frac{\cos{\theta''}}{2\sin{\theta_{n,\alpha}}} \left|
  \sum_{n_2=0}^{N_2(\theta'')} (-i
  e^{\frac{i}{\hbar}S_1^{n,\alpha}})^{n_1(n_2,\theta'') + 1} (i
  e^{\frac{i}{\hbar}S_2})^{n_2} \sum_{k'} A_R(n_2,k') \Omega_R(n_2,k')
\right|^2 
\end{equation}
\end{widetext}
with the  integer-valued functions $N_2$ and $n_1$ given by 
\begin{equation}
N_2(\theta'') = \left[ \frac{W - R_1\sin{\theta_{n,\alpha}} -
    R_1\sin{\theta''}}{2R_2\sin{\theta_2}} \right] \;  ,
\end{equation}
\begin{equation}
\begin{split}
n_1(n_2,\theta'') = \left[ \frac{W - R_1\sin{\theta_{n,\alpha}} -
    R_1\sin{\theta''}}{2R_1\sin{\theta_{n,\alpha}}} \right.
\\
\left. -
  n_2\frac{R_2\sin{\theta_2}}{R_1\sin{\theta_{n,\alpha}}} \right] \; .
\end{split} 
\end{equation}
Note the bound in the integral (\ref{Rnalphafinal}) is $\theta_R = \text{min}(\theta_{n,\alpha},\pi - \theta_{n,\alpha})$ rather than simply $\theta_{n,\alpha}$. This is because as the edge angle $\theta_{n,\alpha}$ exceeds $\pi/2$, trajectories with a final angle $\theta''$ larger than $\pi -\theta_{n,\alpha} $ traverse the Poincar\'e section twice and scatter once more on the interface (see Fig.~\ref{FigIntBounds}). The limiting angle in Eq.~(\ref{Rnalphafinal}) is then the one for which no further scattering on the potential step can take place.
The integral in Eq.\ (\ref{Rnalphafinal}) can be estimated numerically as a function of the interface length $W$ and the tunable field strengths $V_1$, $V_2$ and $B$. 

The transmission $T_{n,\alpha}$ can be calculated following similar steps. For the reader interested in the technical details, a summary of the derivation is presented in Appendix \ref{app:transmission}. The result reads
\begin{widetext}
\begin{equation}
\label{Tnalphafinal}
T_{n,\alpha} = \int_{-\theta_T}^{\theta_T} d \theta''
\frac{R_2\cos{\theta''}}{2R_1\sin{\theta_{n,\alpha}}} \left|
  \sum_{n_2=0}^{N_2(\theta'')} (-i
  e^{\frac{i}{\hbar}S_1^{n,\alpha}})^{n_1(n_2,\theta'')} (i
  e^{\frac{i}{\hbar}S_2})^{n_2} \sum_{k''} A_T(n_2,k'')
  \Omega_T(n_2,k'') \right|^2 
\end{equation}
\end{widetext}
with $\theta_T = \text{min}(\theta_2,\pi - \theta_2)$. The integer valued functions $N_2$ and $n_1$ are given by the formulae 
\begin{equation}
N_2(\theta'') = \left[ \frac{W - R_2\sin{\theta_2} +
    R_2\sin{\theta''}}{2R_2\sin{\theta_2}} \right] \; , 
\end{equation}
\begin{equation}
\begin{split}
n_1(n_2,\theta'') = \left[ \frac{W - R_2\sin{\theta_2} +
    R_2\sin{\theta''}}{2R_1\sin{\theta_{n,\alpha}}} \right.
\\
\left. -
  n_2\frac{R_2\sin{\theta_2}}{R_1\sin{\theta_{n,\alpha}}} \right] \; . 
\end{split}
\end{equation}
Bounds in the integral (\ref{Tnalphafinal}) depend on the sign of $\theta_2 - \pi/2$ for the same reason as for reflected trajectories.

Equations (\ref{Rnalphafinal}) and (\ref{Tnalphafinal}) readily give the total reflection and transmission coefficients, namely,  $R = \sum_{n,\alpha} R_{n,\alpha}$ and $T = \sum_{n,\alpha} T_{n,\alpha}$. Alternatively, these results can be inserted into the Landauer formula,  Eq.~(\ref{LanButt}), giving the conductance. These are, from the technical point of view, the main results of this paper.

\begin{figure}
\begin{center}
\includegraphics[width=0.85\linewidth]{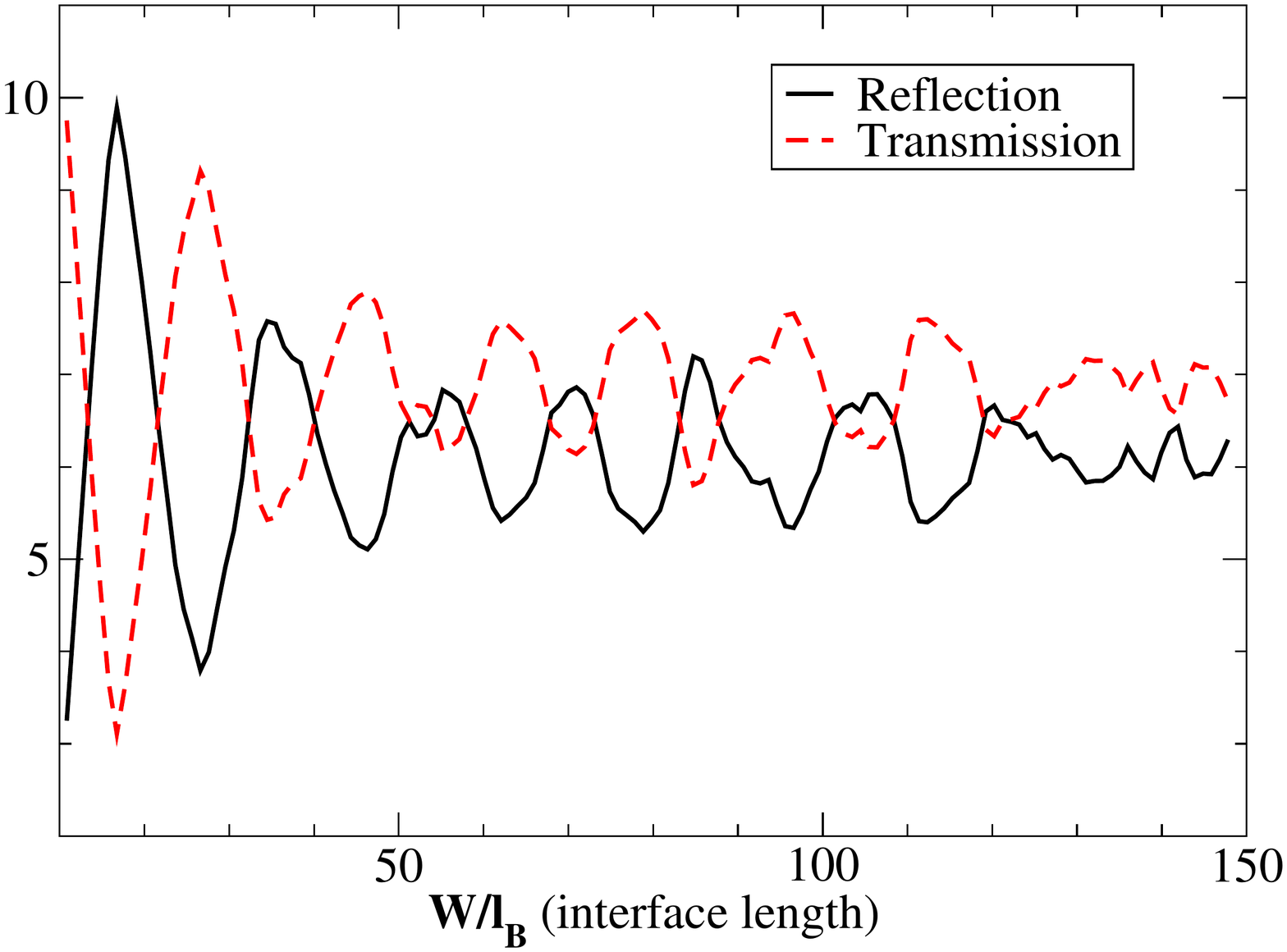}
\includegraphics[width=0.85\linewidth]{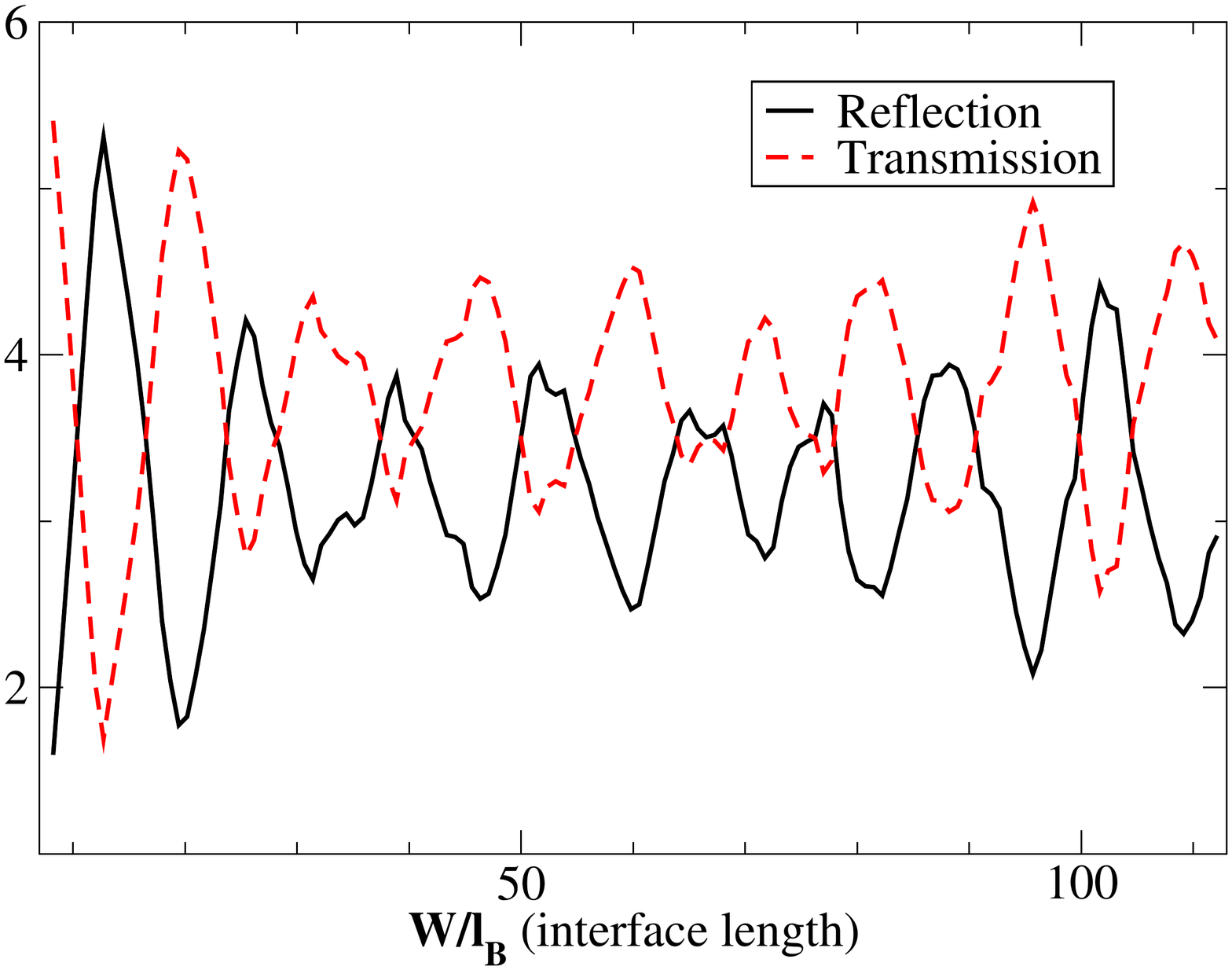}
\caption{(Color online) Total reflection and transmission probabilities 
  as a function of $W/l_B$. {\bf Top}: for filling factors $\nu_1 = 6.37$ 
  and $\nu_2 = 12.15$ in the electron and hole regions respectively. 
  {\bf Bottom}: for filling factors $\nu_1 = 3.70$ and $\nu_2 = 6.99$. 
  The qualitative behavior of individual channel reflection and 
  transmission probabilities is very similar.}
\label{FigJonctPar}
\end{center}
\end{figure}

In Fig.~\ref{FigJonctPar}, we show $R$ and $T$ for a couple of values of the electrostatic potentials $V_1$ and $V_2$. The reflection and transmission coefficients show an oscillating behavior as a function of the interface length $W$.

The overall behavior of $R$ and $T$ is qualitatively similar, but not identical, to the one observed in the symmetric case studied in the previous section. In particular, no saturation of the conductance is observed or predicted. Before discussing these results with greater depth, let us first gain some insight on what is expected classically in the general situation $|V_1| \neq V_2$.


\subsection{Comparison of classical/semiclassical predictions and summary}

\begin{figure}
\begin{center}
\includegraphics[width=0.95\linewidth]{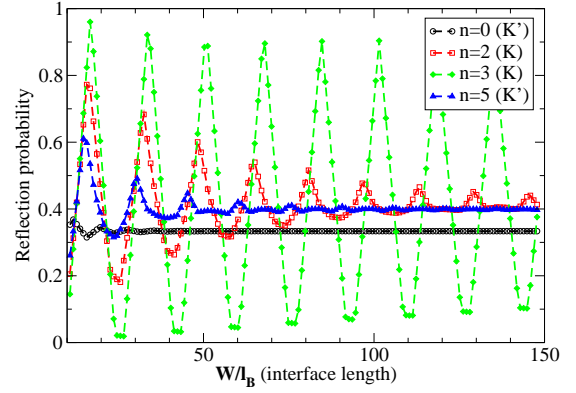}
\caption{(Color online) Classical reflection probabilities of individual channels 
  for a filling factor $\nu_1=6.37$ in the electron region and $\nu_2=12.15$ in the 
  hole region. Each channel is characterized by a Klein transmission probability 
  $T(\theta_{n,\alpha})$: $T(\theta_{0,-}) = 0.45$, $T(\theta_{2,+}) = 0.90$, 
  $T(\theta_{3,+}) = 0.991$, $T(\theta_{5,-}) = 0.78$.  When $T \simeq 1$, classical 
  convergence is masked by large oscillations on the scale of the considered 
  interface length $W$.}
\label{FigReflClass}
\end{center}
\end{figure}

Contrary to the semiclassical probabilities derived in the previous subsection, their classical counterparts converge to an asymptotic value in the limit of a long enough interface. This can be easily shown using the following line of reasoning. Let us call ${\cal P}_1(x)$ and ${\cal P}_2(x)$ the classical probabilities for a charge carrier to be found in region $1$ and region $2$ at a longitudinal distance $x$ from the initial Poincar\'e section. These
probabilities obey equations
\begin{equation}
\left\lbrace
\begin{array}{l}
{\cal P}_1(x) = (1 - T){\cal P}_1(x - L_1) + T{\cal P}_2(x - L_2)
\vspace*{0.2cm}
\\
{\cal P}_2(x) = (1 - T){\cal P}_2(x - L_2) + T{\cal P}_1(x - L_1)
\end{array}
\right.
\end{equation}
whose only asymptotically constant solution as $x \gg L_{1,2}$ is ${\cal P}_1 = {\cal P}_2 = 1/2$. Noticing additionally that charge carriers emerging in region $1$ (respectively region $2$) must do so at a distance smaller than $L_1$ (respectively smaller than $L_2$) from the final Poincar\'e section, one gets, in the asymptotic limit, the classical reflection and transmission probabilities
\begin{equation}
\left\lbrace
\begin{array}{l}
R_{n,\alpha}^{cl} \to L_1/(L_1 + L_2)
\vspace*{0.2cm}
\\
T_{n,\alpha}^{cl} \to L_2/(L_1 + L_2)
\end{array}
\right. \text{as} \; W \to +\infty \; .
\end{equation}
These values are indeed those observed if one plots the classical counterparts of the Fisher-Lee formulae (\ref{Rnalphafinal}) and (\ref{Tnalphafinal}), which can be obtained by considering electrons and holes as non-interfering classical particles. This essentially amounts to replacing probability amplitudes by probabilities and neglecting all phase factors accumulated along the trajectories, giving
\begin{equation}
\begin{split}
R_{n,\alpha}^{cl} = \int_{-\theta_R}^{\theta_R} d \theta''
\frac{\cos{\theta''}}{2\sin{\theta_{n,\alpha}}}
\sum_{n_2=0}^{N_2(\theta'')} (1 - T)^{n_1(n_2,\theta'') + n_2 + 1}
\\
\times \sum_{k'} \left(\frac{T}{1 - T}\right)^{2k'} \Omega_R(n_2,k') \; ,
\end{split} 
\end{equation}
\begin{equation}
\begin{split}
T_{n,\alpha}^{cl} = \int_{-\theta_T}^{\theta_T} d \theta''
\frac{R_2\cos{\theta''}}{2 R_1\sin{\theta_{n,\alpha}}} T
\sum_{n_2=0}^{N_2(\theta'')} (1 - T)^{n_1(n_2,\theta'') + n_2}
\\
\times \sum_{k''} \left(\frac{T}{1 - T}\right)^{2k''} \Omega_T(n_2,k'') 
\end{split}
\end{equation}
with $T = T(\theta_{n,\alpha}) = \sin{\theta_{n,\alpha}}\sin{\theta_2}/\cos^2{((\theta_{n,\alpha} - \theta_2)/2)}$ the Klein transmission probability. These classical formulae are plotted numerically as a function of the interface length $W$ on Fig.~\ref{FigReflClass}.  
As can be seen, some of the probabilities converge rather quickly to the asymptotic values mentioned above, while others show slow convergence, sometimes barely visible on the scale (value of $W$) used. This is of course simply due to the fact that the convergence speed depends on the value of the Klein transmission probability. When the latter approaches unity, the potential barrier becomes transparent for charge carriers which are thus alternatively reflected or transmitted. 

\begin{figure}
\begin{center}
\includegraphics[width=0.9\linewidth]{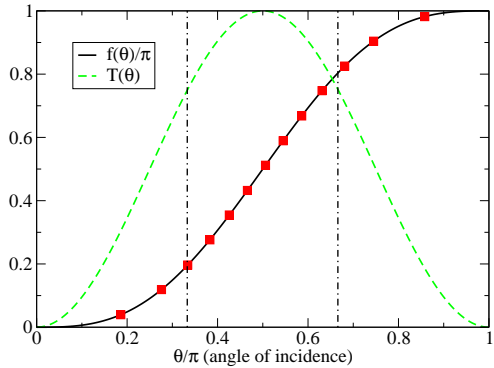}
\caption{(Color online) Functions $f(\theta)/\pi$ (solid, black online) and 
  $T(\theta)$ (dashed, green online). Squares (red online) mark the values of 
  the quantized angles for the filling factor $\nu_1 = 6.37$. As can be clearly 
  seen, most quantized angles sit in the central region $\theta \in [\pi/3,2\pi/3]$, 
  bounded by the vertical dot-dashed lines, and three of them are sufficiently near 
  $\pi/2$ that the semiclassical behavior and the classical one are almost 
  indistinguishable for the interface lengths considered.}
\label{FiggraphS2}
\end{center}
\end{figure}

Quasi-unit Klein transmission probabilities are obtained for angles close to $\pi/2$ which, as illustrated in Fig.~\ref{FiggraphS2}, are more densely sampled when quantizing the dispersion relation (\ref{quantS2}). This behavior is however mainly due to the assumption we made of considering an extremely abrupt potential step on the scale of the magnetic length. Restoring a finite steepness to the $n$-$p$ junction would have the dashed (green online) curve in Fig.~\ref{FiggraphS2} look much more like a sharp peak and considerably reduce the likeliness of having quantized angles with a close to unit Klein transmission probability\,\footnote{Note however that when the filling factor $\nu$ is half-integer, the distribution of quantized angles (\ref{quantS2}) is easily shown to be symmetric with respect to $\pi/2$, which implies the existence of a channel ``sitting" exactly at $\theta=\pi/2$ (since the total number of channels is odd).}.

Coming back to the semiclassical formulae (\ref{Rnalphafinal}) and (\ref{Tnalphafinal}) plotted in Fig.~\ref{FigJonctPar} and as we already pointed out, their behavior is qualitatively very similar to what was observed in the symmetric case plotted in Fig.~\ref{FigScattering}. We lack here the equivalent of the Bloch sphere picture valid in the symmetric case, but we believe nothing fundamentally different is going on here and the physics is hence
essentially the same. The presence of large oscillations in the semiclassical transmissions, as opposed to the classical ones, once again indicates that interferences between trajectories at the potential interface are strong. Concerning the mean values of the probabilities plotted in Fig.~\ref{FigJonctPar}, we found that: {\it (i)} They still differ from the classical limit (as in the symmetric case), and {\it (ii)} They compare poorly with the full mode mixing hypothesis prediction, given by Eq.~(\ref{RMTfmm}). These discrepancies clearly invalidate the possibility of full chaotic mixing in our model, which can be straightforwardly understood by observing that the accessible region of phase space on the hole-doped side of the junction is restricted by Eq.~(\ref{conserv}). A better approximation to the mean semiclassical value can actually be obtained using a mode mixing hypothesis with an effective number of modes in region 2 corresponding to the accessible region of phase space (angles $\theta_2$ in $[-\theta_{crit},\theta_{crit}]$), i.e.
\begin{equation}
\label{renorm}
N_2 \to  N'_2 \stackrel{def}{=} \frac{1}{\hbar}\int_{\pi -
  \theta_{crit}}^{\theta_{crit} - 
  \pi} p_y d y \simeq N_2 \left( 1 - \frac{f(\theta_{crit})}{\pi}
\right) \; .
\end{equation}
Partitioning the current with equal probability in the available edge channels yields a reasonable approximation (about 10\%) to the mean conductance: $\langle G/G_0 \rangle_W \simeq N_1N'_2/(N_1 + N'_2)$. 

Summarizing, in this section we show that, in the semiclassical limit, the general (non-symmetric) case shows large transmission/reflection fluctuations. Furthermore, suppressing these fluctuations by averaging over $W$ would not recover the full mode mixing prediction. A reasonably good approximation of the mean transmission or reflection values can however be obtained from a modified mode mixing hypothesis where only the number of modes corresponding to the accessible phase space is taken into account.

As in the symmetric case, mesoscopic corrections lead unavoidably to quantum fluctuations. Those are clearly non-universal: for certain values of $V_1$ and $V_2$ the magnitude of the fluctuations decreases with increasing $W/l_B$, while for other combinations of $V_1$ and $V_2$ the fluctuations do not seem to depend on $W/l_B$ and no systematic behavior is observed. Notice that we investigate $W/l_B$ values that are comparable with the experimental ones. The variety of fluctuation patterns we observe can be semiclassically explained by means of the quantum interference between different snake-like trajectories.

\section{Model assumptions, limitations, and extensions}
\label{secDisc}

The purpose of this section is to discuss the most important assumptions made in our model. Some of them were introduced for the sake of simplicity and do not introduce limitations to our analysis. Others were necessary to proceed analytically and need further justification.

We begin addressing the geometry of our transport configuration. Recall that instead of the experimental setup of a perpendicular junction, we have settled for the analytically simpler setup of a ``parallel" junction (see the discussion at the beginning of section \ref{secElecAmb}). This issue is addressed in Ref.~[\onlinecite{Carmier11c}] where we demonstrate that, aside for possible diffraction effects at the edge-junction corner, and despite a notable difference in complexity when it comes to exhaustively identifying all trajectories connecting the initial and final Poincar\'e sections, the conductance is very similar in both configurations.

Let us now discuss the effect of different boundary conditions on the conductance. For the magnetic regime, this has already been discussed at the end of section \ref{secMag}, so we shall limit the discussion here to the electric regime. The main effect of choosing zigzag or armchair edges instead of an infinite mass confinement is to modify the quantization condition (\ref{quantS1}) and thereby the values of the quantized angles for the edge channels. The EBK semiclassical quantization procedure was recently applied to the zigzag and armchair cases \cite{Rakyta09} and yields quantization conditions which can be obtained very simply from the infinite mass case by computing the new phase shift $\phi_z$ (or $\phi_a$) acquired by a plane wave scattering on a zigzag (or armchair) edge in graphene. As we have seen however, the conductance of individual channels did not show any special feature dependent on the value of the quantized edge angles $\theta_{n,\alpha}$. All edge channels had a qualitatively similar behavior, their conductance oscillating as a function of $W$. We therefore do not expect that taking zigzag or armchair boundary conditions (which basically amounts to changing the quantized values of the edge angles) will qualitatively alter our predictions.


We now address the effect of replacing our step-like interface by one with a finite steepness $L \ll l_B$. As long as electrons can locally still be approximated as plane waves near the interface, a finite steepness in the potential step essentially sharpens significantly the angular profile of the Klein transmission probability (the dashed green curve in Fig.~\ref{FiggraphS2}) around the angle of perfect transmission. An analytical expression for the Klein transmission probability in this context was derived by Shytov and collaborators \cite{Shytov07}. It is found that the Klein transmission probability is perfect for the incident angle $\theta_\beta = \pi - \cos^{-1}{\beta}$ \footnote{In other words, the transmission remains perfect for normal incidence in the $B = 0$ Lorentz frame, since this is a robust property of graphene electrons in the absence of inter-valley scattering.} and decreases exponentially with $L$ as the angle of incidence is brought away from $\theta_\beta$. A slight asymmetry between Klein transmission probabilities on both sides of the junction is additionally created when $-V_1 \neq V_2$. Nevertheless, the results presented in this paper should  qualitatively hold true as long as $L \ll l_B$ (or equivalently $\beta \ll 1$). 

Let us end our ``tour d'horizon'' by discussing the effect of disorder in the system. Decreasing further the steepness of the barrier favors the charge carriers to dwell longer at the vicinity of the Dirac point which, in most experimental setups, is characterized by the existence of electron-hole puddles \cite{Martin08} combined with a weak screening of electrostatic charges (due to the vanishing density of states). These effects tend to enhance the influence of impurities on the electrons at the interface region \cite{Zhang08,Fogler08}, and possibly drive the system out of the fully coherent ballistic regime that we consider in this paper. Evidently, the inclusion of disorder in our model favors a transition to the regime described by RMT. Indeed, the presence of impurities at the junction interface would randomize the scattering angles and suppress the restriction on the available phase space in region $2$ imposed by Eq.~(\ref{conserv}), bringing the average conductance closer to the full mode mixing value, Eq.\ (\ref{RMTfmm}). We however expect a rather smooth transition from the ballistic non-universal fluctuations we calculate here to the universal ones, ubiquitous in chaotic and disordered systems.

\section{Conclusion}
\label{secConc}

We have studied electronic transport in a graphene $n$-$p$ junction subjected to a strong perpendicular magnetic field. Our main interest in this problem was twofold. Our first goal was to shed some light on the experimentally observed conductance plateaus in this configuration, which still lack a full theoretical explanation. Second, we wanted to confront the full mode mixing hypothesis with a full analysis of a model as consistent as possible with the most physically relevant parameters. The latter was further motivated by the fact that an elementary classical calculation predicts equal reflection and transmission probabilities through a symmetric graphene $n$-$p$ junction in the Hall regime, which raised the question of how much this property would remain true within a semiclassical description.

Concerning our first goal, in distinction to the UCF observed in numerical simulations of disordered $n$-$p$ junctions \cite{Li08,Long08}, we obtain large non-universal transmission fluctuations in ballistic junctions. The combination of these results essentially rules out the possibility of explaining the observations of fractional conductance plateaus within a fully coherent description of the junction. To reconcile experimental and theoretical results,  we believe that a decoherence mechanism suppressing the interference effects is necessary \cite{Abanin07sci}. Among the various possibilities, such as  interactions with electron-hole puddles, electron-phonon and electron-electron interactions, one should find one that is  particularly effective in the vicinity of the junction. For instance, it is plausible that experimental $n$-$p$ junctions are not as abrupt as the ones considered in our model, belonging to an intermediate regime of $\beta \simeq 1$ where a random network of electron-hole puddles \cite{Cheianov07prl} could provide both random current partitioning and decoherence. With the recent advent of hBN substrates \cite{Dean10,Xue11,Mayorov11} which, when intercalated between graphene and SiO$_2$, were shown to significantly increase electrical mobilities and equally reduce charge density inhomogeneities, we hope new experiments can shed more light on the nature of the dephasing mechanism taking place at the junction interface. The effect of inelastic scattering events near the Dirac point certainly also deserves future investigation \cite{Staley08}.

Concerning our second goal, as already stated, we observe that interference effects play a dramatic role for the transport in ballistic $n$-$p$ junctions irrespective of the ratio $W/l_B$. Our expectation of a self-averaging mechanism in the summation over a large number of semiclassical contributing terms was not fulfilled. Surprisingly, in the general case of $V_1 \neq -V_2$, the mean transport quantities (reflection and transmission probabilities) expected from the classical dynamics of the junction differ considerably from their semiclassical (and thus quantum) counterparts. In the case of symmetric junctions, for which a simple and transparent scattering matrix approach can be used, these interference effects can furthermore be described within a Bloch sphere picture, which makes it possible to give a natural interpretation of the discrepancies between classical and semiclassical behaviors. 

Our results suggest the need of further experimental insight to understand the absence of transmission fluctuations in $n$-$p$ junctions at the quantum Hall regime. We believe that the conductance fluctuations predicted here should be observable in junctions for which the decoherence processes are suppressed, particularly for suspended graphene.

\begin{acknowledgments}
We acknowledge a fruitful discussion with Alfredo Ozorio de Almeida. We are also grateful to Oleksii Shevtsov and Xavier Waintal for their help regarding the use of KNIT. This research was supported by the CAPES/COFECUB (project Ph 606/08).
\end{acknowledgments}

\appendix

\section{Calculation of the Maslov index $\mu_J$}
\label{appMaslov}

In this appendix we compute the Maslov indices that appear in Eqs.\ (\ref{propagTnalpha}) and (\ref{propagTnalpha2}). Recall the Green's function Maslov index $\mu_J$ counts the number of caustics on the trajectory $j$, a caustic being a point where $j$ and the neighboring trajectories obtained by an infinitesimal change of the initial momentum intersect each other. Here however it should be noted that, since in the hole region momentum and velocity have an opposite direction, the caustics there should be counted negatively.

Caustics can be found in three different parts of the trajectory: before the first encounter of the trajectory with the interface of the junction; after the last encounter with the interface; and in-between these two points. We shall call respectively $\mu_J'$, $\mu_J''$ and $ \mu_J^c$ the three corresponding contributions to the Maslov index, with $\mu_J = \mu_J' + \mu_J^c + \mu_J''$. We further note $L' = R_1(\sin{\theta_1} - \sin{\theta'})$ the distance between the first encounter and the initial Poincar\'e section, and $L'' = R_i(\sin{\theta''} + \sin{\theta_i})$ ($i= 1$ for reflected trajectories, $i=2$ for transmitted trajectories) the distance between the last encounter and the final Poincar\'e section. Reflected trajectories and transmitted ones will be addressed separately.
\begin{figure}
\begin{center}
\includegraphics[width=0.99\linewidth]{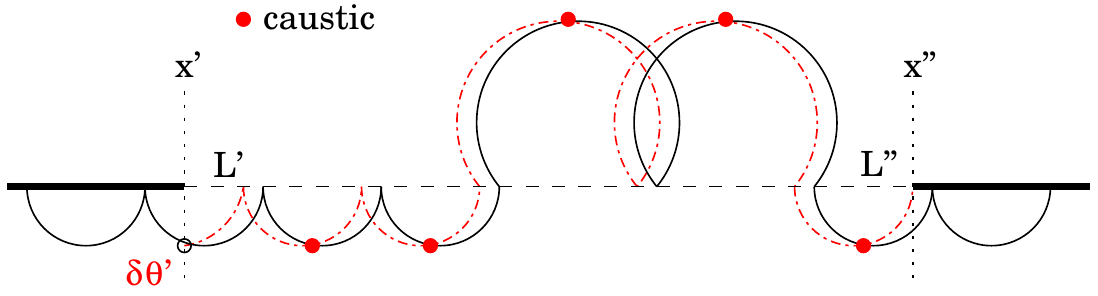}
\includegraphics[width=0.99\linewidth]{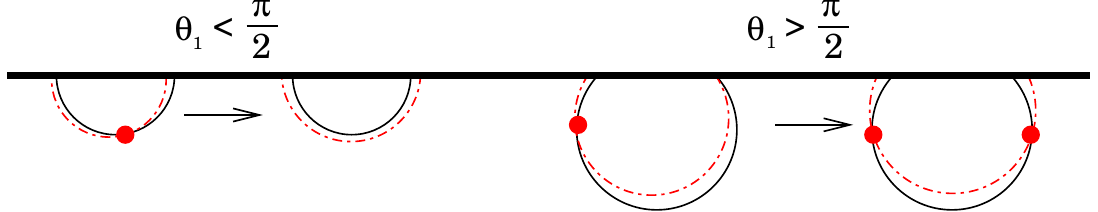}
\caption{(Color online) \textbf{Top}: Caustics along a reflected trajectory 
  with $\sigma_1 < 0$. The rightmost caustic concerns the final portion $L''$ 
  and originates from the fact that $\sigma_2 > 0$. \textbf{Bottom}: Variation 
  of the number of caustics $\mu_J^c$ as $\sigma_1$ becomes positive. If 
  $\theta_1 < \pi/2$, the transitional excursion contains no caustic, while if 
  $\theta_1 > \pi/2$ the transitional excursion contains two caustics.}
\label{FigMaslovI}
\end{center}
\end{figure}

\subsection{Reflected trajectories}

Let us introduce the following indices
\begin{equation}
\label{indicesigma}
\sigma_1 = \text{sign}\left( \left. \frac{\partial L'}{\partial
      \theta'} \right|_{y'} \left. \frac{\partial L''}{\partial
      \theta'} \right|_{y'} \right) \; ,
\end{equation}
\begin{equation}
\sigma_2 = \text{sign}\left( \left. \frac{\partial L''}{\partial 
      \theta'} \right|_{y'} \left. \frac{\partial y''}{\partial 
      \theta'} \right|_{y'} \right) \; .
\end{equation}
Noting that keeping $y' = R_1 \cos{\theta_1} - R_1 \cos{\theta'}$ constant leads to $\sin{\theta'} d\theta' = \sin{\theta_1} d \theta_1$, $\sigma_1$ and $\sigma_2$ can be computed from
\begin{equation}
\left. \frac{\partial L'}{\partial \theta'} \right|_{y'} = R_1
\frac{\sin{(\theta' - \theta_1)}}{\sin{\theta_1}} \; , 
\end{equation}
\begin{equation}
\begin{split}
R_1^{-1}\left. \frac{\partial L''}{\partial \theta'} \right|_{y'} =
  \cos{\theta'} - (2n_1 +
  1)\cos{\theta_1}\frac{\sin{\theta'}}{\sin{\theta_1}}
\\
-2n_2\cos{\theta_2}\frac{\sin{\theta'}}{\sin{\theta_2}} \; , 
\end{split}
\end{equation}
\begin{equation}
\begin{split}
\left( R_1^{-1} \left. \frac{\partial y''}{\partial \theta'} \right|_{y'}
+\sin{\theta'} \right) (\tan{\theta''})^{-1} = \cos{\theta'}
\\
 - 2(n_1 + 1)\cos{\theta_1}\frac{\sin{\theta'}}{\sin{\theta_1}}
 - 2n_2\cos{\theta_2}\frac{\sin{\theta'}}{\sin{\theta_2}} \; . 
\end{split}
\end{equation}

Looking at Fig.~\ref{FigMaslovI} it is clear that if $\sigma_1 < 0$ each excursion contains exactly one caustic, and the contribution $\mu_J^c$ from the central part of the trajectory is exactly $n_1 - n_2 $.  If $\sigma_1 > 0$, however, one of the excursions will be in the configuration schematized on the lower part of Fig.~\ref{FigMaslovI}, and will contain either two or zero caustics. This transitional excursion can take place either in region 1 or in region 2. In the former case the number of caustics in the transitional excursion is zero if $\theta_1 < \pi/2$ and two if $\theta_1 > \pi/2$, and vice versa in the latter case. Since however caustics are counted with an opposite sign in regions 1 and 2, this leaves the contribution $\mu_J^c$ unaffected.

In the same way the number of caustics in the last excursion is one if $\sigma_2 > 0$, and zero otherwise. Finally, a caustic can be found in the first excursion if $\theta_1 - \theta' > \pi$. All this can be summarized by
\begin{equation}
\label{muJresult}
\left\lbrace
\begin{array}{l}
\mu''_J = \Theta(\sigma_2)
\vspace*{0.2cm}
\\
\mu_J^c = n_1 - n_2 + \text{sign}(\theta_1 - \pi/2)\Theta(\sigma_1)
\vspace*{0.2cm}
\\
\mu'_J = \Theta(\theta_1 - \theta' - \pi)
\end{array}
\right.
\end{equation}
with $\Theta$ the Heaviside step function.

\subsection{Transmitted trajectories}

This time, the final position is $y'' = R_2 \cos{\theta''} - R_2 \cos{\theta_2}$. Still working with $y'$ fixed leads to the same expressions as above for $({\partial L'}/{\partial \theta'})_{y'}$ and $({\partial L''}/{\partial \theta'})_{y'}$, while the variation of the final position can be shown to read
\begin{equation}
\begin{split}
\left( R_1^{-1} \left. \frac{\partial y''}{\partial \theta'} \right|_{y'} 
+\sin{\theta'} \right) (\tan{\theta''})^{-1} = \cos{\theta'} 
\\
- (2n_1 +
    1)\cos{\theta_1}\frac{\sin{\theta'}}{\sin{\theta_1}}
- (2n_2 +
    1)\cos{\theta_2}\frac{\sin{\theta'}}{\sin{\theta_2}} \; . 
\end{split}
\end{equation}
Defining the same indices as for the reflected trajectories, the same expression is obtained for the number of caustics in the central part of the trajectory. Concerning the extremal portions, a caustic can be found in $L'$ if  $\theta_1 - \theta' - \pi > 0$ (as before) and one in $L''$ this time if $\sigma_2 < 0$.

\subsection{Total Maslov index}

For semiclassical expressions such as Eq.~(\ref{interR}), it is the sum $\mu_J+\mu_S$ which is relevant rather than $\mu_J$ alone. Although we will not provide a formal proof of this, it can be seen (and it is easily checked numerically) that 
\begin{equation}
\mu_J + \mu_{S} =
\mu(\nu') - \mu(\nu'') + n_1 - n_2 + 1 \; .
\end{equation}
We make use of this equality in section~\ref{secElecGen}.

\section{Calculation of the prefactor $J\;{\partial^2_{y'} S}$}
\label{appPref}

In this appendix, we compute the prefactor obtained in Eq.~(\ref{interR}) once integral (\ref{calTnalpha}) has been evaluated in the stationary phase approximation. It reads 
\begin{equation}
\label{AnxC1}
J_j(y'_s) \left. \frac{\partial^2 (S_j + S_{n,\alpha})}{\partial y'^2}
\right|_{y''}(y'_s) \; . 
\end{equation}
Recall $S_j$ is the action of the Green's function, $S_{n,\alpha}$ that of the mode, $J_j$ the Green's function prefactor and $y'_s$ the stationary phase point. Using basic properties of the action we have 
\begin{equation}
J_j(y'_s) = \dot{x'} \dot{x''} \left( -\frac{\partial^2 S_j}{\partial
    y'' \partial y'} \right)^{-1}(y'_s) = \dot{x'} \dot{x''}
\left. \frac{\partial y''}{\partial p'_y} \right|_{y'}(y'_s) 
\end{equation}
with $\dot{x'} = v_F\cos{\theta_{n,\alpha}(y'_s)}$ and $\dot{x''} = v_F\cos{\theta_{n,\alpha}(y'')}$, and
\begin{equation}
\left. \frac{\partial^2 (S_{n,\alpha} + S_j)}{\partial y'^2}
\right|_{y''} = \left. \frac{\partial (p_{y'}^{n,\alpha} -
    p'_y)}{\partial y'} \right|_{y''} \; . 
\end{equation}
Now, for a variation within the manifold on which the mode $(n,\alpha)$ is built, $y'' = y''(y',p'_y(y',(n,\alpha)))$, and thus 
\begin{equation}
d y'' = \left( \left. \frac{\partial y''}{\partial y'} \right|_{p'_y}  +
\left. \frac{\partial y''}{\partial p'_y} \right|_{y'}
\left. \frac{\partial p'_y}{\partial y'} \right|_{n,\alpha} \right) d
y' \; .  
\end{equation}
Using then that $\left. \left( {\partial y''}/{\partial y'} \right) \right|_{p'_y} = - \left. \left( {\partial p'_y}/{\partial y'} \right) \right|_{y''} \left. \left( {\partial y''}/{\partial p'_y} \right) \right|_{y'}$ we obtain
\begin{equation}
\left( -\frac{\partial^2 S_j}{\partial y'' \partial y'} \right)^{-1}
\left. \frac{\partial^2 (S_j + S_{n,\alpha})}{\partial y'^2}
\right|_{y''} = \frac{d y''}{d y'} \; , 
\end{equation}
which expresses the usual ratio between measures $d y''$ and $d y'$ on the corresponding Poincar\'e sections. Making use of the identity $W = (n_1 + 1)L_1 + n_2L_2 + R_1(\sin{\theta''} - \sin{\theta'})$, it can be easily evaluated as 
\begin{equation}
\begin{split}
\frac{d y''}{d y'} = \left. \frac{\partial y''}{\partial \theta''}
\right|_{n,\alpha} \left. \frac{\partial \theta''}{\partial \theta'}
\right|_{n,\alpha} \left. \frac{\partial \theta'}{\partial y'}
\right|_{n,\alpha} = R_1\sin{\theta_{n,\alpha}(y'')}
\\
\times
\left(\frac{\cos{\theta_{n,\alpha}(y')}}{\cos{\theta_{n,\alpha}(y'')}}\right)
(R_1\sin{\theta_{n,\alpha}(y')})^{-1} 
\end{split}
\end{equation}
and expression (\ref{AnxC1}) hence takes the final form
\begin{equation}
\begin{split}
J_j(y'_s) \left. \frac{\partial^2 (S_j + S_{n,\alpha})}{\partial y'^2}
\right|_{y''}(y'_s) = \left(v_F \cos{\theta_{n,\alpha}(y'_s)}\right)^2
\\
\times
\frac{\sin{\theta_{n,\alpha}(y'')}}{\sin{\theta_{n,\alpha}(y'_s)}} \; .
\end{split} 
\end{equation}

\section{Main steps for the derivation of $T_{n,\alpha}$}
\label{app:transmission}

Here we describe how to deal with the case of transmitted trajectories (for which $\epsilon({\bf r''}) < 0$) and obtain the transmission $T_{n,\alpha}$. We discuss only the few main steps of the calculation that differ from the derivation of $R_{n,\alpha}$, presented in the main text. 

This time, the matrix structure of the integral that appears in Eq.~(\ref{calTnalpha}) reads
\begin{equation}
\begin{split}
V_j^{-}({\bf r''}) V_j^{+\dagger}({\bf r'}) \sigma_x
\left( \begin{array}{c} e^{-i\frac{\nu'}{2}\theta_{n,\alpha}(y')} \\
    e^{i\frac{\nu'}{2}\theta_{n,\alpha}(y')} \end{array} \right) =
e^{-i\theta_j'/2} 
\\
\times \cos{\left(\frac{\nu'\theta_{n,\alpha}(y') +
    \theta_j'}{2}\right)} \left( \begin{array}{c} e^{-i\theta_j''} \\
    -1 \end{array} \right) \; . 
\end{split}
\end{equation}
The Green's function projector $V_j^-({\bf r''})V_j^+({\bf r'})$ selects those trajectories which connect the initial Poincar\'e section in the positive (electron) eigenspace and the final Poincar\'e section in the negative (hole) eigenspace. Applying the stationary phase approximation to integral (\ref{calTnalpha}) yields the same condition as the one obtained in the main text for the reflection, and the action of the Green's function reads $S_j(y'_s) = \hbar k_x^{n,\alpha}(x'' - x') + (n_1 + 1/2)S_1^{n,\alpha} + (n_2 + 1/2)S_2 + \nu'' S(y'') - \nu' S_{n,\alpha}(y'_s)$. The Maslov index is computed as in the case of reflected trajectories (details can be found in appendix \ref{appMaslov}).

By putting together these elements, the integral (\ref{calTnalpha}) can be evaluated as
\begin{widetext}
\begin{equation}
\label{propagTnalpha2}
\begin{split}
{\cal T}_{n,\alpha}({\bf r''}) =
\frac{C_{n,\alpha}e^{ik_x^{n,\alpha}x''}}{\sqrt{|\sin{\theta''}|}}
\sum_{\nu'' = \pm 1} e^{\frac{i}{\hbar}\nu'' S(y'') +
  i\frac{\pi}{2}\mu(\nu'')} e^{-i\nu''\theta''}
\left( \begin{array}{c} e^{-i\nu\theta''/2} \\ -
    e^{i\nu\theta''/2} \end{array} \right)
e^{\frac{i}{\hbar}(S_1^{n,\alpha} + S_2)/2} 
\\
\times -it_1 \sum_{n_2=0}^{N_2} (-i
e^{\frac{i}{\hbar}S_1^{n,\alpha}})^{n_1} (i
e^{\frac{i}{\hbar}S_2})^{n_2} \sum_{k''} A_T(n_2,k'')
\Omega_T(n_2,k'') \; . 
\end{split}
\end{equation}
\end{widetext}
As expected (and contrary to the case of reflected trajectories), the incident mode once propagated to the point ${\bf r''}$ is no longer quantized. This, however, turns out to be irrelevant when it comes to computing the conductance making use of Eq.~(\ref{Tnalpha}). Switching from position to angular coordinates, the transmission probability of channel $n$ polarized in valley $\alpha$ gives Eq.\ \eqref{Tnalphafinal}.

\bibliography{snake1}
\bibliographystyle{prsty}

\end{document}